\documentclass[draftclsnofoot,onecolumn,12pt]{IEEEtran}
\usepackage{graphicx,cite,amssymb,amsmath,psfrag,subfigure}
\usepackage{graphicx}
\usepackage{amssymb}
\usepackage{amsmath}
\usepackage{cite}
\usepackage{subfigure}
\usepackage{mathrsfs}
\usepackage[displaymath,mathlines]{lineno}
\usepackage{color}
\usepackage{tabulary}
\usepackage{multirow}
\usepackage{url}
\usepackage{bm}

\newtheorem{theorem}{Theorem}

\newtheorem{corollary}{Corollary}
\newtheorem{proposition}{Proposition}

\newtheorem{remark}{Remark}

\newcommand{\mbs}[1]{\bm{#1}}
\newcommand{\mat}[1]{{\uppercase{\mbs{#1}}}}
\newcommand{\Id}{\mat{\mathrm{I}}}

\def\nn{\nonumber}
\renewcommand{\H}{{\scriptscriptstyle\mathsf{H}}}

\DeclareMathAlphabet{\mathpzc}{OT1}{pzc}{m}{it}

\DeclareMathOperator{\E}{\mathbb{E}} \DeclareMathOperator{\M}

\newcommand{\EX}[1]{\E\left\{{#1}\right\}}
\newcommand{\EXs}[2]{\E_{{#1}}\left\{{#2}\right\}}

\newcommand{\PDF}[2]{p_{{#1}}\left({#2}\right)}
\newcommand{\CG}[2]{\mathcal{CN}\left({#1},{#2}\right)}

\newcommand{\B}[1]{{\pmb{#1}}}

\newcommand{\Pu}{p_{r}}

\def\nn{\nonumber}
\renewcommand{\H}{{\scriptscriptstyle\mathsf{H}}}

\DeclareMathAlphabet{\mathpzc}{OT1}{pzc}{m}{it}

\makeatletter
\def\@setsize#1#2#3#4{
    \@nomath#1
    \let\@currsize#1
    \baselineskip #2
    \baselineskip \baselinestretch\baselineskip
    \parskip \baselinestretch\parskip
    \setbox\strutbox \hbox{
        \vrule height.7\baselineskip
            depth.3\baselineskip
            width\z@}
    \skip\footins \baselinestretch\skip\footins
    \normalbaselineskip\baselineskip#3#4}
\makeatother

\makeatletter
\newcommand{\setstretch}[1]{
    \def\baselinestretch{#1}%
    \@currsize
    }
\makeatother



\def\BibTeX{{\rm B\kern-.05em{\sc i\kern-.025em b}\kern-.08em
    T\kern-.1667em\lower.7ex\hbox{E}\kern-.125emX}}

\newcounter{eqncnt}

\newcounter{eqnback}

\begin{document}

\title{
    Uplink Performance  of Conventional and Massive MIMO Cellular Systems with Delayed CSIT}
\author{
       Anastasios K. Papazafeiropoulos, Hien Quoc  Ngo, and
        Tharm Ratnarajah
\thanks{Parts of this work were presented at the 2014 IEEE International Symposium on Personal, Indoor and Mobile Radio Communications (PIMRC)~\cite{Papazafeiropoulos4}.}

\thanks{
        A. K. Papazafeiropoulos is with Communications and Signal Processing Group, Imperial College London, London, U.K. (email:
a.papazafeiropoulos@imperial.ac.uk). }
\thanks{
        H.~Q.\ Ngo is with the Department of Electrical
    Engineering (ISY), Link\"{o}ping University, 581 83 Link\"{o}ping,
    Sweden
        (email: nqhien@isy.liu.se).
}
\thanks{
T. Ratnarajah is with Institute for Digital Communications
(IDCoM), University of Edinburgh, Edinburgh, U.K. (email:
t.ratnarajah@ed.ac.uk).}

 \thanks{This research was supported by a Marie Curie Intra European Fellowship and HARP project
 within the 7th European Community Framework Programme for Research of the European Commission under grant agreements no. [330806],
 IAWICOM and no. [318489], HARP.}

}


\maketitle

\begin{abstract}
Massive multiple-input multiple-output (MIMO) networks, where
the base stations (BSs) are equipped with large number of antennas
and serve a number of users simultaneously, are very promising,
but suffer from pilot contamination. Despite its importance, delayed channel state information (CSI) due to user
mobility, being another degrading factor, lacks investigation in
the literature. Hence, we consider an uplink model, where each BS
applies zero-forcing decoder, accounting for both effects, but
with the focal point on the relative users' movement with regard
to the BS antennas. In this setting, analytical closed-form
expressions for the sum-rate with finite number of BS antennas,
and the asymptotic limits with infinite number of BS antennas
epitomize the main contributions. In particular, the probability
density function of the signal-to-interference-plus-noise ratio
 and the ergodic sum-rate are derived for any finite number
of antennas. Insights of the impact of the arising Doppler shift
due to user mobility into the low signal-to-noise ratio regime as
well as the outage probability are obtained. Moreover, asymptotic
analysis performance results in terms of infinitely increasing
number of antennas, power, and both numbers of antennas and users
(while their ratio is fixed) are provided. The numerical results
demonstrate the performance loss in various Doppler shifts. An
interesting observation is that massive MIMO is favorable even
in time-varying channel conditions.
\end{abstract}


\section{Introduction} \label{Sec:Introduction}

The rapidly increasing demand for wireless connectivity and
throughput is one of the motivations for the continuous evolution
of cellular networks \cite{Huawei,gesbert}. Very large
multiple-input multiple-output (MIMO) has been identified as a new
promising breakthrough technique aiming at achieving higher area
throughput in wireless networks \cite{Marzetta,Rusek,Ngo_Energy}.
Its origin is found in \cite{Marzetta}, and it has been given many
alternative names such as massive MIMO, hyper-MIMO, and
full-dimension MIMO systems. In the typical envisioned
architecture, each base station (BS) with an array of hundreds or
even thousands antennas, exploiting the key idea of multi-user
MIMO (MU-MIMO), serves tens or hundreds of single-antenna users
simultaneously in the same frequency band, respectively, under
coherent processing. This difference in the number of BS antennas
$N$ and the number of users $K$ per cell provides unprecedented
spatial degrees of freedom that leads to high signal gains,
allowing at the same time low-complexity linear signal processing
techniques and avoiding inter-user interference due to the (near)
orthogonality between the channels.

On a similar note, zero-forcing (ZF) processing is regarded as a
low-complexity alternative of  maximum-likelihood multiuser
detector and ``dirty paper coding'' \cite{Wang}, especially, when
the BSs are equipped with massive antenna arrays. A lot of
research has been conducted on single-cell systems with ZF
receivers \cite{Caire}, but the main current interest has shifted
to practical multi-cell scenarios, where pilot contamination
degrades the system performance~\cite{Marzetta,Jose}.

Despite that the theory of massive MIMO has been now well
established (see \cite{Marzetta} and references therein), an
important question that has been overlooked is how the performance
of massive MIMO topology is affected by the relative movement
of users. This scenario is of high practical importance, in urban
environments, where users move rapidly within a geographical area.
The main challenge in time-varying environments is to perform
robust channel estimation, when the propagation channel changes
over time. The dynamic channel behavior was modeled in terms of a
stationary ergodic Gauss-Markov block fading channel
model~\cite{Truong,Papazafeiropoulos1,Papazafeiropoulos2}, where
an autoregressive model was combined with the Jakes'
autocorrelation function that captures the time variation of the
channel. Motivated by the above observation, this paper explores
the robustness offered by massive MIMO against the practical
setting of user mobility that results to delayed and degraded
channel state information (CSI) at the BS, and thus, imperfect
CSI. Such consideration is notably important because it can
provide the quantification of the performance loss in various
Doppler shifts.

A limited effort for studying the time variation of the channel
due to the relative movement of users has been conducted
in~\cite{Truong}, where the authors provided deterministic
equivalents (DEs)\footnote{The deterministic equivalents are
deterministic tight approximations of functionals of random
matrices of finite size. Note that these approximations are
asymptotically accurate as the matrix dimensions grow to infinity,
but can be precise for small dimensions.} for the maximal ratio
combining (MRC) receiver in the uplink and the matched filter (MF)
in the downlink transmission. Fortunately, this analysis was
extended in~\cite{Papazafeiropoulos1,Papazafeiropoulos2} by
deriving DEs for the minimum mean-square error (MMSE) receiver
(uplink) and regularized zero-forcing (downlink) and by making a
comparison regarding their performance. In this paper, we consider
a generalized uplink massive MIMO system. Based on the
aforementioned literature, we propose a tractable model that
encompasses ZF receivers and describes the impact of user mobility
in a cellular system with BSs having conventional and very large
number of antennas, or even large number of both antennas and
users, which stands in contrast to the previous works. The
following are the main contributions of this paper:
\begin{itemize}
\item Contrary to \cite{Performance_hien}, we consider more
practical settings where the channel is imperfectly estimated at
the BS. The effects of pilot contamination and time variation of
the channels are taken in to account. The extension is not
straightforward because apart of the development of the model, the
mathematical manipulations are hampered. Apart of this, the
results are contributory and novel.

\item We derive the probability density function (PDF) of the
signal-to-interference-plus-noise ratio (SINR) and the
corresponding ergodic sum-rate for any finite number of antennas
in closed forms. For the sake of completeness, the link of these
results with previous known results is mentioned. Furthermore, a
simpler and more tractable lower bound for the achievable uplink
rate is derived.

\item We elaborate on the low signal-to-noise ratio (SNR) regime,
in order to get additional insights into the impact of Doppler
shift. In particular, we study the behaviors of the minimum
normalized energy per information bit to reliably convey any
positive rate and the wideband slope.

\item We present a simple expression for the outage probability,
being an important metric in quasi-static models.

\item We investigate the asymptotic performance presented by very
large MIMO ($N\to \infty$) as well as large MIMO in terms of DEs
($N, K\to \infty$). This analysis aims at providing accurate
approximation results that replace the need for lengthy Monte
Carlo simulations.
\end{itemize}

Note that, although all the results incur significant mathematical
challenges, they can be easily evaluated. Nevertheless, the
purpose of DEs is to provide the deterministic tight
approximations, in order to avoid lengthy Monte-Carlo simulations.

\emph{Notation:} We use boldface lowercase and uppercase letters
to denote vectors and matrices, respectively. The notation
$(\cdot)^\H$ stands for the conjugate transpose, and $\|\cdot\|$
denotes the Euclidean norm of a vector, while $(\cdot)^{\dagger}$
denotes the pseudo-inverse of a matrix. We use the notation $x
\overset{\tt d}{\sim} y$ to imply that $x$ and $y$ have the same
distribution. Finally, we use $\B{z} \sim
\CG{\mathbf{0}}{\B{\Sigma}}$ to denote a circularly symmetric
complex Gaussian vector $\B{z}$ with zero mean and covariance
matrix $\B{\Sigma}$.

\section{System Model} \label{sec: system}

Consider a cellular network which has $L$ cells. Each cell
includes one $N$-antenna BS and $K$ single-antenna users. We
consider the uplink transmission. The model is based on the
assumptions that: i) $N\geq K$, and ii) all users in $L$ cells
share the same time-frequency resource.
 Moreover, we assume that the
channels are frequency flat and they vary from symbol to symbol,
while during the symbol period they are considered constant due to
the channel aging impact \cite{Truong} (we will discuss about the
channel aging model later). The channel vector $\B
g_{lik}[n]\in\mathbb{C}^{N\times 1}$ between the $k$th user in the
$i$th cell and the $l$th BS at the $n$th symbol undergoes
independent small-scale fading and large-scale fading. More
precisely, $\B g_{lik}[n]$ is modelled as
\begin{align}
\B g_{lik}[n]= \sqrt{\beta_{lik}} \B h_{lik}[n],
\end{align}
where $\beta_{lik}$ represents  large-scale fading, and $\B
h_{lik}\in\mathbb{C}^{N\times 1}\sim \CG{\B{0}}{\Id_N}$ is the
small-scale fading vector between the $l$th BS and the $k$th user
in the $i$th cell.

Let $\sqrt{p_{r}} \B{x}_i[n]\in \mathbb{C}^{K\times 1}$ be the
zero-mean stochastic data signal vector of $K$ users in the $i$th
cell at time instance $n$ ($p_{r}>0$ is the average transmitted
power of each user, and $\B{x}_i[n]\sim  \CG{\B{0}}{\Id_N}$). Then, the $N\times1$ received signal vector
at the $l$th BS is
\begin{align}
    \B{y}_l[n]
    =
        \sqrt{p_{r}}
        \sum_{i=1}^{L} \B{G}_{li}[n]  \B{x}_i[n]   +  \B{z}_l[n],~~~ l=1, 2, ...,
        L,\label{eq MU-MIMO 1}
\end{align}
where $\B G_{li}[n]\triangleq \left[\B g_{li1}[n],\ldots, \B
g_{liK}[n]\right]\in \mathbb{C}^{N\times K}$ denotes the channel
matrix between the $K$ users in the $i$th cell and the $l$th BS,
and $\B{z}_l[n]\sim \CG{\B{0}}{\Id_N}$ is additive white
Gaussian noise (AWGN) vector at the $l$th BS.

To coherently detect the signals transmitted from the $K$ users in
the $l$th cell, the BS needs CSI knowledge. Conventionally, the
$l$th BS can estimate the channel via uplink training.
We assume that the channel remains
constant during the training phase \cite{Truong}. In general, this
assumption is not practical, but it yields a simple model which
enables us to analyze the system performance and to obtain initial
insights on the impact of channel aging. Furthermore, the impact
of channel aging can be absorbed in the channel estimation error.

During the training phase, in each cell, $K$ users are assigned
$K$ orthogonal pilot sequences of length $\tau$ symbols ($\tau\geq
K$). Owing to the limitation of the coherence interval, the pilot
sequences have to be reused from cell to cell. We assume that all
$L$ cells use the same set of orthogonal pilot sequences. As a
result, the pilot contamination occurs \cite{Marzetta,Rusek}.
Denote by $\B{\Psi}\in\mathbb{C}^{K\times \tau}$, ($\tau \geq K$),
be the pilot matrix transmitted from the $K$ users in each cell,
where the $k$th row of $\B{\Psi}$ is the pilot sequence assigned
for the $k$th user. The matrix $\B{\Psi}$ satisfies $\B{\Psi}
\B{\Psi}^\H=\Id_K$. Then, the $N \times \tau$ received pilot
signal at BS $l$ is
\begin{align}\label{eq MU-MIMO training}
\B{Y}^{\mathrm{tr}}_l[n]
    =        \sqrt{p_{\mathrm{tr}}}
        \sum_{i=1}^{L} \B{G}_{li}[n]\B {\Psi}   +  \B{Z}^{\mathrm{tr}}_l[n],~~~ l=1, 2, ..., L,
\end{align}
where  the subscript $()^\mathrm{tr}$ implies the  uplink training
stage, ${p_{\mathrm{tr}}}\triangleq \tau {p_{r}}$,  and
$\B{Z}_l^{\mathrm{tr}}[n]\in \mathbb{C}^{N\times \tau}$ is
spatially AWGN at BS $l$ during the training phase. We assume the
the elements of $\B{Z}^{\mathrm{tr}}_l[n]$ are i.i.d.\ $\CG{0}{1}$
random variables (RVs). The MMSE channel estimate of ${\B
g}_{lik}[n]$ is given by \cite{Truong}
\begin{align} \label{eq:MMSEchannelEstimate}
\hat{\B g}_{lik}[n] = & \beta_{lik} \B Q_{lk} \left( \sum_{j =1}^L
\B g_{ljk}[n] + \frac{1}{\sqrt{p_{\mathrm{tr}}}}\tilde{\B
z}^{\mathrm{tr}}_{lk}[n] \right)\!\!,
\end{align}
where $\B Q_{lk}\! \triangleq \!\left(\frac{1}{p_\mathrm{tr}}\!
+\! \sum_{i=1}^L \beta_{lik}\right)^{\!-1}\!\Id_N$, and $\tilde{\B
z}^{\mathrm{tr}}_{lk}[n] \!\sim\! \mathcal{CN}\!\left( \B
0,\Id_{N} \right)$ represents the additive noise.

With MMSE channel estimation, the channel estimate and the channel
estimation error are uncorrelated. Thus, $\B g_{lik}[n]$ can be
rewritten as:
\begin{align}
\B g_{lik}[n] = \hat{\B g}_{lik}[n] + \tilde{\B
g}_{lik}[n],\label{eq:MMSEorthogonality}
\end{align}
where  $\tilde{\B g}_{lik}[n] \sim \CG{\B 0}{\left(
\beta_{lik}-\hat{\beta}_{lik} \right)\B{\mathrm{ I}}_N}$ and
 $\hat{\B g}_{lik}[n] \sim \CG{\B 0}{\hat{\beta}_{lik}}$, with
$\hat{\beta}_{lik}=\frac{\beta^{2}_{lik}}{\sum_{j=1}^{L}\beta_{ljk}+1/p_{\mathrm{tr}}}$,
are the independent channel estimate and channel estimation error,
respectively. Note that   $\beta_{lik}$, $\hat{\beta}_{lik}$, and
$\B Q_{lk}$ are independent of $n$ $\forall l$, $i$,  and $k$.

Besides pilot contamination, in any common propagation scenario, a
relative movement takes place between the antennas and the
scatterers that degrades more channel's  performance. Under these
circumstances, the channel is time-varying and needs to be modeled
by the famous Gauss-Markov block fading model, which is basically
an autoregressive model of certain order that incorporate
two-dimensional isotropic scattering (Jakes model). More
specifically, our analysis achieves to relate the current channel
state with its past samples. For the sake of analytical
simplicity, we consider the following simplified autoregressive
model of order $1$~\cite{Truong}
\begin{align}\label{eq:aut}
 \B g_{lik}[n]=\alpha {\B g}_{lik}[n-1]+\B e_{lik}[n],
\end{align}
where  ${\B g}_{lik}[n-1]$ and $\B e_{lik}[n]\sim
\CG{\B{0}}{\left( 1-\alpha^2 \right)\beta_{lik}\B{\mathrm{ I}}_N}$
are uncorrelated, denoting the channel at the previous symbol
duration and the stationary Gaussian channel error vector due to
the time variation of the channel, respectively. Note that
$\alpha\!=\!\mathrm{J}_{0}\left( 2 \pi f_{D} T_{s} \right)$, where
$\mathrm{J}_{0}(\cdot)$ is the zeroth-order Bessel function of the
first kind, $f_{D}$ and $T_{s}$ are the maximum Doppler shift and
the channel sampling period. Basically, $\alpha$, which is assumed known at the BS, corresponds to
the temporal correlation parameter that describes the isotropic
scattering according to the Jakes' model. In particular, the
maximum Doppler shift $f_{D}$ equals $f_{D}=\frac{v f_{c}}{c}$,
where $v$ (in m/s) is the relative velocity of the user,
$c=3\times10^{8}$m/s is the speed of light, and $f_{c}$ is the
carrier frequency.

Substituting \eqref{eq:MMSEorthogonality} into \eqref{eq:aut}, we
obtain a model which combines both effects of channel estimation
error and channel aging as follows:
\begin{align}
 \B g_{lik}[n]&=\alpha {\B g}_{lik}[n-1]+\B e_{lik}[n]\nonumber\\
&=\alpha \hat{\B g}_{lik}[n-1]+ \tilde{\B
e}_{lik}[n],\label{eq:MMSEchannelEstimate}
\end{align}
where $\hat{\B g}_{lik}[n-1]$ and $\tilde{\B
{e}}_{lik}[n]\triangleq \alpha \tilde{\B g}_{lik}[n-1]+\B
e_{lik}[n]\sim \CG{\B{0}}{\left(
\beta_{lik}-\alpha^{2}\hat{\beta}_{lik} \right)\B{\mathrm{ I}}_N}$
are mutually independent.  More concretely, we define $\hat{\B
G}_{li}[n]\triangleq\left[\hat{\B g}_{li1}[n],\ldots, \hat{\B
g}_{liK}[n]\right]\in \mathbb{C}^{N\times K}$ and $\tilde{\B
E}_{li}\triangleq\left[\tilde{\B e}_{li1}[n],\ldots, \tilde{\B
e}_{liK}[n]\right]\in \mathbb{C}^{N\times K}$ as the combined
channel matrices from all users in cell $i$ to BS $l$.
In particular, $\hat{\B G}_{li}[n]$ can be expressed as
\cite{Performance_hien}:
\begin{align}\label{interchannelEstimated}
 \hat{\B G}_{li}[n]= \hat{\B G}_{ll}[n] \B {{D}}_{i},
\end{align}
where $\B
{{D}}_{i}=\mathrm{diag}\big\{\frac{{\beta}_{li1}}{{\beta}_{ll1}},\frac{{\beta}_{li2}}{{\beta}_{ll2}},\ldots,\frac{{\beta}_{liK}}{{\beta}_{llK}}
\big\}$.

Making use of~\eqref{eq:MMSEchannelEstimate}, we can rewrite the
received signal $ \B{y}_l[n]$ at the $l$th BS $\left(
l\in\left[1,L  \right] \right)$ as
\begin{small}
\begin{align}\label{eq MU-MIMO 1}
    \B{y}_l[n]
   \! =\!
        \alpha \sqrt{p_{r}}\!
        \sum_{i=1}^{L}\! \!\B{\hat{G}}_{li}[n\!-\!1]  \B{x}_i[n]  \! + \! \sqrt{p_{r}}\! \sum_{i=1}^{L} \!\!\B{\tilde{E}}_{li}[n]  \B{x}_i[n] \! + \!\B{z}_l[n].
\end{align}
\end{small}
Moreover, we assume that the $l$th BS uses the ZF technique to
detect the signals transmitted from $K$ users in its cells. With
ZF, the received signal vector $\B{y}_l[n]$ is pre-multiplied with
$\alpha^{-1}\hat{\B{G}}_{ll}^\dagger[n-1]$, where
$\hat{\B{G}}_{ll}^\dagger[n-1]$ is a $K\times N$ matrix
representing the pseudo-inverse of $\hat{\B{G}}_{ll}[n-1]$:
\begin{align}
 \label{eq MU-MIMO 2}
\! &\!\B{r}_l[n]
    =
       \sqrt{p_{r}}\B{x}_l[n]+
         \sqrt{p_{r}} \sum_{i\ne l}^{L} \hat{\B{G}}_{ll}^\dagger[n-1] \B{\hat{G}}_{li}[n-1]  \B{x}_i[n]            \nn\\
  \!  \!\!\! \! \!  \!\! &\!+\!  \alpha^{\!-\!1}\!\sqrt{p_{r}} \sum_{i=1}^{L}\! \hat{\B{G}}_{ll}^\dagger [n\!-\!1] \B{\tilde{E}}_{li}[n]  \B{x}_i[n] \!+\!\alpha^{\!-\!1}\hat{\B{G}}_{ll}^\dagger[n\!-\!1] \B{z}_l[n],
\end{align}

\setcounter{eqnback}{\value{equation}} \setcounter{equation}{11}
\begin{figure*}[!t]
\begin{align}\label{eq: SINR}
    \!  \!  \!  \!   \gamma_k
   \! =\!
        \frac{
            \alpha^2\Pu
            }{
            \alpha^2\Pu
            \sum_{i \neq l}^{L}\!
                \left\|\!
                    \left[\hat{\B{G}}_{ll}^{\dagger}[n\!-\!1]\right]_k
                   \! \hat{\B{G}}_{li}[n\!-\!1]
                \right\|^2
            \!+\!
        \Pu
            \sum_{i = l}^{L}\!
                \left\|\!
                    \left[\hat{\B{G}}_{ll}^{\dagger}[n\!-\!1]\right]_k
                   \! \tilde{\B{E}}_{li}[n]
                \!\right\|^2\!+\!
                 \left\|\!
                    \left[\hat{\B{G}}_{ll}^{\dagger}[n\!-\!1]\right]_k
                                   \! \right\|^2
            }.
\end{align}
\hrulefill
\end{figure*}
\setcounter{eqncnt}{\value{equation}}
\setcounter{equation}{\value{eqnback}}
Then, the $k$th element of $\B{r}_l[n]$ is used to detect the
signal transmitted from the $k$th user. The post-processed
received signal corresponding to the $k$th user is
\begin{flalign}
 \label{eq MU-MIMO 2}
  \!\!  &\!\!\!{r}_{lk}[n]
    =
         \sqrt{p_{r}}{x}_{lk}[n]\!+\!
         \sqrt{p_{r}}\! \sum_{i\ne l}^{L}\! \left[\hat{\B{G}}_{ll}^\dagger[n-1]\right]_{k}\! \B{\hat{G}}_{li}[n\!-\!1]  \B{x}_i[n]  \nn\\
          \!\!\!\! &\!\!\!+ \! \frac{1}{\alpha}\!\sqrt{p_{r}} \!\!\sum_{i=1}^{L} \!\left[\hat{\B{G}}_{ll}^\dagger[n\!-\!1]\right]_{k}\!\!\B{\tilde{E}}_{li}[n] \B{x}_i[n]  \!+\!\frac{1}{\alpha}\!\!\left[\hat{\B{G}}_{ll}^\dagger[n\!-\!1]\right]_{k}\! \B{z}_l[n],   \end{flalign}
where the notation $\left[\B A\right]_{k}$ refers to the $k$th row
of matrix $\B A$, and ${x}_{lk}[n]$ is the $k$th element of
$\B{x}_{l}[n]$, i.e., it is the transmit signal from the $k$th
user in the $l$th cell at the $n$th time slot. Treating~\eqref{eq
MU-MIMO 2} as a single-input single-output (SISO) system, we
obtain the SINR of the transmission from the $k$th user in the
$l$th cell to its BS in \eqref{eq: SINR} shown at the top of the
page. The SINR is obtained under
the assumption that the $l$th BS knows the denominator value of
\eqref{eq: SINR}. This assumption is reasonable since this value
is just a scalar (which can be estimated).

\section{Achievable Uplink Sum Rate} \label{Sec:Rate}
In this section, we provide the sum-rate analysis for finite and
infinite number of BS antennas taking into account the
aforementioned effects.

\subsection{Finite-$N$ Analysis}\label{Sec:Rate_CF}

\begin{proposition}\label{Prop 1}
The uplink SINR of transmission between the $k$th user in the
$l$th cell to its BS, under the delayed channels, is distributed
as
\setcounter{equation}{12}\begin{align} \label{eq Prop1 1}
    \gamma_k
    \mathop \sim \limits^{\tt d}
                \frac{
                    \alpha^2 p_{r} X_k[n-1]
                    }{
                    \alpha^2p_{r} C X_k[n-1] +p_{r} Y_k[n]+ 1
                    },
\end{align}
where $X_k$ and $Y_k$ are independent random variables (RVs) whose
PDFs are, respectively, given by
\begin{align} \label{eq PDF1 1}
   \!\!\PDF{X_k}{x}
   &\!=\!
        \frac{
            e^{-x/\hat{ \beta}_{llk}}
            }{
            \left(N-K\right)!
           \hat{ \beta}_{llk}
            }
        \left(
            \frac{
                x
                }{
               \hat{ \beta}_{llk}
                }
        \right)^{N-K}, ~ x \geq 0
    \\
    \!\!\PDF{Y_k}{y}
    &\!=\!\!
        \sum_{p=1}^{\varrho\left(\! \mathbf{\mathcal{A}}_k\!\right)}\!
        \sum_{q=1}^{\tau_p \left(\! \mathbf{\mathcal{A}}_k\!\right)} \!\!
            \!\!\mathcal{X}_{p,q} \left(\! \mathbf{\mathcal{A}}_k\!\right)
           \! \frac{
                \mu_{k,p}^{-q}
            }{
                \left(
                    q\!-\!1
                \right)
                !
            }
            y^{q-1}
            e^{\frac{-y}{\mu_{k,p}}}
      ,
         ~
        y \geq 0,\label{eq PDF1 3}
\end{align}
where $\mathbf{\mathcal{A}}_k\triangleq \mathrm{diag} \left(
\tilde{\B{D}}_{l1},\ldots,\tilde{\B{D}}_{lL} \right)\in
\mathbb{C}^{K L\times K L} $  with $\tilde{\B D}_{li}$ a $K \times
K$ diagonal matrix having elements $\left[\tilde{\B
D}_{li}\right]_{kk}=\left( \beta_{lik}-\alpha^{2}\hat{\beta}_{lik}
\right)$, as well as $\varrho\left( \mathbf{\mathcal{A}}_k\right)$
denotes the numbers of distinct diagonal elements of
$\mathbf{\mathcal{A}}_k$. Similarly, $\mu_{k,1}, \mu_{k,2}, ...,
\mu_{k,\varrho\left( \mathbf{\mathcal{A}}_k\right)}$ are the
associated distinct diagonal elements in decreasing order and
$\tau_p \left( \mathbf{\mathcal{A}}_k\right)$ are the
multiplicities of $\mu_{k,p}$, while $\mathcal{X}_{p,q} \left(
\mathbf{\mathcal{A}}_k\right)$ is the $\left(p,q\right)$th
characteristic coefficients of $\mathbf{\mathcal{A}}_k$, as
defined in \cite[Definition~4]{SW:08:IT}. Regarding $C$,  it is a
deterministic constant: $C \triangleq \sum_{i \ne
l}^{L}\left(\frac{\beta_{lik}}{\beta_{llk}} \right)^2$.
\begin{proof}
Division of each term of \eqref{eq: SINR} by $\left\|
\left[\hat{\B{G}}_{ll}^{\dagger}[n-1]\right]_k \right\|^2$ leads
to
\begin{align} \label{eq Rate 1}
 \!\! \!\!\!  \!\!\! \!\gamma_k
    \!=\!
        \frac{
            \alpha^2\Pu
            \left\| \left[\hat{\B{G}}_{ll}^{\dagger}[n-1]\right]_k
          \right\|^{-2}
            }{
           \alpha^2 \Pu C
            \left\| \left[\hat{\B{G}}_{ll}^{\dagger}[n\!-\!1]\right]_k
            \right\|^{-2}
            \!\!+\!
        \Pu \sum_{i =1}^{L}
                \left\|
                    \hat{\B{Y}}_i[n]
                \right\|^2
\!\!+\!
            1
            },
\end{align}
where $ C \triangleq \sum_{i \ne
l}^{L}\left\|\!\left[\hat{\B{G}}_{ll}^{\dagger}[n-1]\right]_k
\hat{\B{G}}_{li}[n-1] \!\right\|\mathop = \limits^{\tt
\eqref{interchannelEstimated}}\sum_{i \ne
l}^{L}\left(\frac{\beta_{lik}}{\beta_{llk}} \right)^2$ and
$\hat{\B{Y}}_i[n] \triangleq
\frac{\left[\hat{\B{G}}_{ll}^{\dagger}[n-1]\right]_k
\tilde{\B{E}}_{li}[n]}{\left\|\left[\hat{\B{G}}_{ll}^{\dagger}[n-1]\right]_k \right\|}$.\\
Since $\left\| \left[\hat{\B{G}}_{ll}^{\dagger}[n-1]\right]_k
\right\|^2 = \left[\left(\hat{\B{G}}_{ll}^\H[n-1] \hat{\B{G}}_{ll}
[n-1]\right)^{-1} \right]_{kk},$ $\left\|
\left[\B{G}_{ll}^{\dagger}[n-1]\right]_k \right\|^{-2}$  has an
Erlang distribution with shape parameter $N-K+1$ and scale
parameter $\hat{\beta}_{llk}$ \cite{GHP:02:CL}. Then,
\begin{align} \label{eq PDF1 1a}
    \left\| \left[\hat{\B{G}}_{ll}^{\dagger}[n-1]\right]_k \right\|^{-2}
    \mathop \sim \limits^{\tt d}
    X_k[n-1].
\end{align}
Furthermore, conditioned on
$\left[\hat{\B{G}}_{ll}^\dagger[n-1]\right]_k$, $\hat{\B{Y}}_i[n]$
is a zero-mean complex Gaussian vector with covariance matrix
$\tilde{\B D} _{li}$ which is independent of
$\left[\hat{\B{G}}_{ll}^\dagger[n-1]\right]_k$. Thus,
$\hat{\B{Y}}_i[n]\sim \CG{\B{0}}{\tilde{\B{D}}_{li}}$, independent
of $\left[\hat{\B{G}}_{ll}^\dagger[n-1]\right]_k$. Hence,
$\sum_{i=1}^{L}\left\| \hat{\B{Y}}_i[n] \right\|^2$ is the sum of
$K L$ statistically independent but not necessarily identically
distributed exponential RVs. According
to~\cite[Theorem~2]{BSW:07:WCOM}, we obtain
\begin{align}
 \sum_{i= l}^{L}
                \left\|
                    \hat{\B{Y}}_i[n]
                \right\|^2
   \overset{\tt d}{\sim}
    Y_k[n].\label{eq PDF1 y}
\end{align}
Combining \eqref{eq Rate 1}--\eqref{eq PDF1 y}, we deduce
\eqref{eq Prop1 1}.
\end{proof}
\end{proposition}

\begin{remark}\label{remark 1}
In the general case, the PDF of the uplink SINR~\eqref{eq Prop1 1}
accounts for both the effects of pilot contamination and Doppler
shift. More specifically, the time variation of the channel
decreases both the desired and interference signal powers by a
factor of~$\alpha^2$ with comparison to the \emph{zero}th Doppler
shift case, thus, degrading the SINR.
\end{remark}

 \begin{remark}
Increasing the relative velocity, the SINR presents ripples with
peak and zero points following the behaviour of the
$\mathrm{J}_{0}(\cdot)$ Bessel function. In the marginal case of
$\alpha=1$, i.e., when there is no relative movement of the
user,~\eqref{eq Prop1 1} expresses the downgrade of the system
only to pilot contamination. Especially, if we assume very long
training intervals (orthogonal pilot sequences) and no time
variation, which is not practical in common scenarios with large
number of antennas and moving users, our result coincides
with~\cite[Eq.~(6)]{Mathaiou:ZF_receivers}. At the other end, high
velocity meaning $\alpha \to 0$ leads to zero SINR.
 \end{remark}

 \begin{corollary}
Consider the high uplink power regime. We have
\begin{align} \label{pu limit}
 \gamma_k
    \mathop \sim \limits^{\tt d}
            \frac{
            \alpha^2  X_k[n-1]
            }{
            \alpha^2 C X_k[n-1]
                 +Y_k[n]}, \quad \text{as} ~ p_{r} \rightarrow \infty.
\end{align}
\setcounter{eqnback}{\value{equation}} \setcounter{equation}{19}
\begin{figure*}[!t]
\begin{align}
  \!  \!    \mathcal{J}_{m,n}\left( a,b,\alpha \right)
    &\triangleq
        \sum_{r=0}^{m}
           \!   \binom{m}{r} \! 
            \left(-b\right)^{m-r} \! 
        \left[
            \sum_{s=0}^{n+r}
            \frac{
                \left(n+r \right)^s
                b^{n+r-s}
                }{
                \alpha^{s+1} a^{m-s}
                }
            \mathrm{Ei} \left(\!-b\right)
            -
            \frac{
                \left(n+r \right)^{n+r}
                e^{\alpha b/a}
                }{
                \alpha^{n+r+1}
                a^{m-n-r}
                }
            \mathrm{Ei} \left(\!-\frac{\alpha b}{a}-b \!\right)
        \right.
    \nonumber
    \\
    & \hspace{3.8 cm}
        \left.
        +
        \frac{e^{-b}}{\alpha }
        \sum_{s=0}^{n+r-1}
        \sum_{u=0}^{n+r-s-1}
        \frac{
            u!
            \left(n+r\right)^s \binom{n+r-s-1}{u}
            b^{n+r-s-u-1}
            }{
            \alpha^s
            a^{m-s}
            \left(\alpha/a + 1 \right)^{s+1}
            }
        \right] \!  \! .\label{I Func}
\end{align}
\hrulefill
\end{figure*}
\setcounter{eqncnt}{\value{equation}}
\setcounter{equation}{\value{eqnback}}

This corollary brings  an important insight on the system
performance, when $p_{r}$ increases asymptotically. As seen
in~\eqref{pu limit}, there is a finite SINR ceiling due to the
simultaneous increment of the desired signal power as well as of
the interference and channel estimation error powers.
\end{corollary}

Having obtained the PDF of the SINR, and by defining the function
$\mathcal{J}_{m,n}\left( a,b,\alpha \right)$ as in \eqref{I Func}
shown at the top of the previous page, where
$\mathrm{Ei}\left(\cdot\right)$ denotes the exponential integral
function \cite[Eq.~(8.211.1)]{GR:07:Book}, we first obtain the
exact $R_{lk}\left( p_{r},\alpha \right)$ and a simpler lower
bound $R_{L}\left( p_{r},\alpha \right)$ as follows:

\begin{theorem} \label{Theo 1}
The uplink ergodic achievable rate  of transmission between the
$k$th user in the $l$th cell to its BS for any finite number of
antennas, under delayed channels, is given by
\setcounter{equation}{20}\begin{align}
   \label{Rate prop}
\!\!R_{lk}\!\left( p_{r},\alpha \right)
    \!=\!\!\!
\sum_{p=1}^{\varrho\left(\! \mathbf{\mathcal{A}}_k\!\right)}\!
        \sum_{q=1}^{\tau_p \left(\! \mathbf{\mathcal{A}}_k\!\right)}\!
           \!\!\! \frac{
                 \mathcal{X}_{p,q} \left(\! \mathbf{\mathcal{A}}_k\!\right)
               \mu_{k,p}^{-q} \log_2 e
            }{
                \left(
                    q\!-\!1
                \right)
        !\!
                \left(N\!-\!K\right)!
                \hat{\beta}_{llk}^{N\!-\!K\!+\!1}\!
            }    \left(  \mathcal{I}_{1}\!-\!\mathcal{I}_{2} \right)\!,
\end{align}
where $\mathcal{I}_{1}$ and $\mathcal{I}_{2}$ are given by
\eqref{eq Prop2 1a} and \eqref{eq Prop2 1b} shown at the top of
next page, and where $U\left(\cdot,\cdot,\cdot \right)$ is the
confluent hypergeometric function of the second kind
\cite[Eq.~(9.210.2)]{GR:07:Book}.
\begin{proof}
See Appendix~\ref{sec:proof:prop rate1}.
\end{proof}
\end{theorem}

\setcounter{eqnback}{\value{equation}} \setcounter{equation}{21}
\begin{figure*}[!t]
\begin{align} \label{eq Prop2 1a}
      \mathcal{I}_{1}
    &\!\triangleq\! \sum_{t=0}^{N-K}\left[
            - e^{\frac{1}{\hat{\beta}_{llk} \alpha^2 p_{r} \left( C+1 \right)}}
            \mathcal{J}_{q\!-\!1, N\!-\!K\!-\!t}\left(\!\frac{1}{\hat{\beta}_{llk}\alpha^{2}\left( C+1 \right)},\frac{1}{\hat{\beta}_{llk} \alpha^2 p_{r} \left( C+1 \right)},\frac{1}{\mu_{k,p}}\!-\!\frac{1}{\hat{\beta}_{llk}\alpha^{2}\left( C+1 \right)}\! \right)
        \right.\nn \\
                &+
                \sum_{u=1}^{N-K-t}
                \frac{\left(u-1 \right)! \left(-1\right)^{u} p_{r}^{-q}}{\left( {\hat{\beta}_{llk}\alpha^2 p_{r} \left( C+1 \right)}\right)^{N-K-t-u} }
                  \left.
        \Gamma\left(q\right)
                U\left(q, q+1+N-K-t-u, \frac{1}{\mu_{k,p}p_{r} }\right)
            \right]\\
                   \mathcal{I}_{2}
    &\!\triangleq\! \! \sum_{t=0}^{N-K}\left[
            - e^{\frac{1}{\hat{\beta}_{llk} \alpha^2 p_{r} C}}
            \mathcal{J}_{q\!-\!1, N\!-\!K\!-\!t}\left(\!\frac{1}{\hat{\beta}_{llk}\alpha^{2}C},\frac{1}{\hat{\beta}_{llk} \alpha^2 p_{r} C},\frac{1}{\mu_{k,p}}\!-\!\frac{1}{\hat{\beta}_{llk}\alpha^{2}C}\! \right)
        \right.\nn\\
        &        +
                \sum_{u=1}^{N-K-t}
                \frac{\left(u-1 \right)! \left(-1\right)^{u} p_{r}^{-q}}{\left( {\hat{\beta}_{llk}\alpha^2 p_{r} C}\right)^{N-K-t-u} }
         \left.
                 \   \Gamma\left(q\right)
                U\left(q, q+1+N-K-t-u, \frac{1}{\mu_{k,p}p_{r} }\right)
            \right]\label{eq Prop2 1b},
\end{align}
\hrulefill
\end{figure*}
\setcounter{eqncnt}{\value{equation}}
\setcounter{equation}{\value{eqnback}}

In the case that all diagonal elements of $\mathbf{\mathcal{A}}_k$
are distinct, we have $\varrho\left( \mathbf{\mathcal{A}}_k\right)
= KL$, $\tau_p \left( \mathbf{\mathcal{A}}_k\right)=1$, and
$\mathcal{X}_{p,1} \left( \mathbf{\mathcal{A}}_k\right)
    =    \prod_{q=1, q \neq p}^{KL}
    \left(
        1
        -
        \frac{\mu_{k,q}}{\mu_{k,p}}
    \right)^{-1}$. The uplink rate becomes
\setcounter{equation}{23}\begin{align} \label{eq Prop2 1}
  R_{lk}\!\left(\! p_{r},\alpha \!\right)    &\!=\!
         \sum_{p=1}^{KL}
    \sum_{t=0}^{N-K}\!
             \frac{
                \prod_{q=1, q \neq p}^{KL}
    \left(\!
        1
        \!-\!
        \frac{\mu_{k,q}}{\mu_{k,p}}
    \!\right)^{-1}\!\!\!\log_2 e
                            }{
                \left(N\!-\!K\!-\!t\right)! (-1)^{N-K-t}\mu_{k,p}
                            }\left( \bar{\mathcal{I}}_1\!-\!\bar{\mathcal{I}}_2 \right),
               \end{align}
where $\bar{\mathcal{I}}_{1}$ and $\bar{\mathcal{I}}_{2}$ are
given by \eqref{eq Prop2 1aa} and \eqref{eq Prop2 1bb} shown at
the top of next page. Note that, we have used the identity
$U\left(1, b, c \right)=e^{x}x^{1-b}\Gamma\left( b-1,x \right)$
\cite[Eq. (07.33.03.0014.01)]{Wolfram} to obtain \eqref{eq Prop2
1}.

\setcounter{eqnback}{\value{equation}} \setcounter{equation}{24}
\begin{figure*}[!t]
\begin{align} \label{eq Prop2 1aa}
      \bar{\mathcal{I}}_{1}
    &\!=\!\sum_{t=0}^{N-K}\left[
            - e^{\frac{1}{\hat{\beta}_{llk} \alpha^2 p_{r} \left( C+1 \right)}}
            \mathcal{J}_{0, N\!-\!K\!-\!t}\left(\!\frac{1}{\hat{\beta}_{llk}\alpha^{2}\left( C+1 \right)},\frac{1}{\hat{\beta}_{llk} \alpha^2 p_{r} \left( C+1 \right)},\frac{1}{\mu_{k,p}}\!-\!\frac{1}{\hat{\beta}_{llk}\alpha^{2}\left( C+1 \right)}\! \right)
        \right.\nn\\
                &+
                \sum_{u=1}^{N-K-t}
                \frac{\left(u-1 \right)! \left(-1\right)^{u} }{\left( {\hat{\beta}_{llk}\alpha^2  \left( C+1 \right)}\right)^{N-K-t-u} }e^{\frac{1}{\mu_{k,p}p_{r}}}\mu_{k,p}^{N+1-K-t-u}\Gamma\left(N+1-K-t-u, \frac{1}{\mu_{k,p}p_{r} }\right)\Biggl. \Biggr]\\
      \bar{\mathcal{I}}_{2}
    &\!=\! \! \sum_{t=0}^{N-K}\left[
            - e^{\frac{1}{\hat{\beta}_{llk} \alpha^2 p_{r} C}}
            \mathcal{J}_{1, N\!-\!K\!-\!t}\left(\!\frac{1}{\hat{\beta}_{llk}\alpha^{2}C},\frac{1}{\hat{\beta}_{llk} \alpha^2 p_{r} C},\frac{1}{\mu_{k,p}}\!-\!\frac{1}{\hat{\beta}_{llk}\alpha^{2}C}\! \right)
        \right.\nn\\
        &        +
                \sum_{u=1}^{N-K-t}
                \frac{\left(u-1 \right)! \left(-1\right)^{u} }{\left( {\hat{\beta}_{llk}\alpha^2  C}\right)^{N-K-t-u} }
         \left.
                    e^{\frac{1}{\mu_{k,p}p_{r}}}\mu_{k,p}^{N+1-K-t-u}\Gamma\left(N+1-K-t-u, \frac{1}{\mu_{k,p}p_{r} }\right)
            \right].\label{eq Prop2 1bb}
\end{align}
\hrulefill
\end{figure*}
\setcounter{eqncnt}{\value{equation}}
\setcounter{equation}{\value{eqnback}}

\setcounter{eqnback}{\value{equation}} \setcounter{equation}{28}
\begin{figure*}[!t]
\begin{align}\label{eq:out}
\begin{small}
 \displaystyle P_{{\tt out}}\!\left( \gamma_{{\tt th}}\right)
 \!=\!
 \left\{\!\!\!
   \begin{array}{l}
     1, \quad \text{if $\gamma_{{\tt th}} \geq 1/C$} \\
      1\!-\!e^{-\frac{\gamma_{{\tt th}}}{\hat{\beta}_{llk}\left( \alpha^2 p_{r}-\alpha^{2} p_{r} C \gamma_{{\tt th}} \right)}}\!
\sum\limits_{p=1}^{\varrho\left( \mathbf{\mathcal{A}}_k\right)}\!
\sum\limits_{q=1}^{\tau_p \left( \mathbf{\mathcal{A}}_k\right)}\!
 \sum\limits_{t=0}^{N-K} \!\!
  \sum\limits_{s=0}^{t}\!\! \binom{t}{s}
 \mathcal{X}_{p,q} \!\left(
 \mathbf{\mathcal{A}}_k\!\right) \frac{\mu_{k,p}^{-q}}{\left(q\!-\!1\right)} \Gamma\left( s\!+\!q \right)\!\!\left( \hat{\beta}_{llk}\left( \alpha^{2}\!- \!\alpha^{2} C \gamma_{{\tt th}}\right) \!\!\right)^{s+q},  \quad \text{if $\gamma_{{\tt th}} < 1/C$}.\\
   \end{array}
 \right.
 \end{small}
\end{align}
\hrulefill
\end{figure*}
\setcounter{eqncnt}{\value{equation}}
\setcounter{equation}{\value{eqnback}}

\begin{proposition}\label{Prop BoundRate 1}
The uplink ergodic rate  from the $k$th user in the $l$th cell to
its BS, considering delayed channels, can be presented by a
certain lower bound $ R_{L}\left( p_{r},\alpha \right)$:
\setcounter{equation}{26}\begin{align}\label{eq: LB}
 \!\! &R_{lk}\left( p_{r},\alpha \right)  \!\geq\! R_{L}\left( p_{r},\alpha
 \right)\nonumber\\
&\triangleq \log_2\!\!\left(\!\!1\!+ \!\frac{1}{C+\frac{1}{\left(
N-K \right)\alpha^{2} \hat{
\beta}_{llk}}\!\left(\sum\limits_{i=1}^{L}\sum\limits_{k=1}^{K}\!\!
\left(\! \beta_{lik}\!-\!\alpha^{2}\hat{\beta}_{lik}
\!\right)\!+\!\frac{1}{p_{r}}\! \right)}\!\!\right)\!.
\end{align}
\begin{proof}
See Appendix~\ref{propBound Rate}.
\end{proof}
\end{proposition}

\subsubsection{Outage Probability}
Bearing in mind that we investigate a block fading model, the
study of the outage probability is of crucial interest. Basically,
it defines the probability that the instantaneous SINR
$\gamma_{k}$ falls below a given threshold value $\gamma_{{\tt
th}}$:
\begin{align}
 P_{{\tt out}}\left( \gamma_{{\tt th}}\right)
 &= {\tt Pr}\left( \gamma_k\leq\gamma_{{\tt th}}\right).
\end{align}
\begin{theorem}\label{theo 2}
The outage probability of transmission from the $k$th user in the
$l$th cell to its BS is given by \eqref{eq:out}, shown at the top
of the page.
\begin{proof}
See Appendix~\ref{proof:Prop out}.
\end{proof}
\end{theorem}

\subsection{Characterization in the Low-SNR Regime}
Even though  Theorem~\ref{Theo 1} renders possible the exact
derivation of the uplink sum-rate, it appears deficient to provide
an insightful dependence on the various parameters such as the
number of BS antennas and the transmit power. On that account, the
study of the low power cornerstone, i.e., the low-SNR regime, is
of great significance. There is no reason to consider the high-SNR
regime, because in this regime an important metric such as the
high-SNR slope $\mathcal{S}_{\infty}=  \lim_{p_{r} \rightarrow 0}
\frac{R_{lk}\left( p_{r}, \alpha \right)}{\mathrm{log}_2 p_{r}}$
\cite{Tulino} is zero due to the finite sum-rate, as shown in
\eqref{pu limit}.

\subsubsection{Low-SNR Regime}
In case of low-SNR, it is possible to represent the rate by means
of second-order Taylor approximation as
\setcounter{equation}{29}\begin{align} R_{lk}\left( p_{r}, \alpha
\right)=\dot{R}_{lk}\left( 0, \alpha
\right)p_{r}+\ddot{R}_{lk}\left( 0, \alpha
\right)\frac{p_{r}^{2}}{2}+o\left( p_{r}^{2} \right),
\end{align}
where $\dot{R}_{lk}\left( p_{r}, \alpha \right)$ and
$\ddot{R}_{lk}\left( p_{r}, \alpha \right)$ denote the first and
second derivatives of $R_{lk}\left( p_{r}, \alpha \right)$ with
respect to SNR $p_{r}$. In fact, these parameters enable us to
examine the energy efficiency in the regime of low-SNR by means of
two key element parameters, namely the minimum transmit energy per
information bit, $\frac{E_{b}}{N_{0}{_{\mathrm{min}}}}$, and the
wideband slope $S_{0}$~\cite{Verdu}. Especially, we have
\begin{align}\label{eq: minimum energy}
 \frac{E_{b}}{N_{0}{_{\mathrm{min}}}}&=\lim_{p_{r} \rightarrow 0}\frac{p_{r}}{R_{lk}\left( p_{r}, \alpha \right)}=\frac{1}{\dot{R}_{lk}\left( 0, \alpha \right)},\\
 S_{0}&=-\frac{2\left[ \dot{R}_{lk}\left( 0, \alpha \right)\right]^{2}}{\ddot{R}_{lk}\left( 0, \alpha \right)}\mathrm{ln}2.\label{eq: wideband_slope}
\end{align}

\begin{theorem}\label{Theo 3}
In the low-SNR regime, the uplink sum-rate betweem the $k$th user
in the $l$th cell to its BS in a multi-cell system, assuming
delayed channels, can be captured by the minimum transmit energy
per information bit, $\frac{E_{b}}{N_{0}{_{\mathrm{min}}}}$, and
the wideband slope $S_{0}$, respectively, expressed by
 \begin{align}
  &\frac{E_{b}}{N_{0}{_{\mathrm{min}}}}=\frac{\mathrm{ln}2}{\alpha^2\left( N-K+1 \right)\hat{\beta}_{llk}}\\
&S_{0}\!= \!\frac{-2\!\left( N\!-\!K\!+\!1 \right)/\left(
N\!-\!K\!+\!2 \right)}{\alpha^{4} \!+\!2 \alpha^{2} C \!\left(\!
N\!-\!K\!+\!3 \!\right)\!+\!\frac{2}{N-K+2}
\!\sum\limits_{p=1}^{\varrho\left(\!
\mathbf{\mathcal{A}}_k\!\right)}
      \!  \sum\limits_{q=1}^{\tau_p \left(\! \mathbf{\mathcal{A}}_k\!\right)} \!
           \! \frac{
                \mathcal{X}_{p,q}\! \left(\! \mathbf{\mathcal{A}}_k\!\right)\mu_{k,p}^{-q}q
            }{
                \hat{ \beta}_{llk}\left(
                    q-1
                \right)
                !
            }}.
            \end{align}
\begin{proof}
See Appendix~\ref{sec:low snr}.
\end{proof}
\end{theorem}

\subsection{Large Antenna Limit Analysis}
In this section, we consider the large system limit by accounting
for  specific assumptions. Assuming constant transmit power
$p_{r}$ and : i) the number of BS antennas $N$ grows infinitely
large, while $K$ is fixed, ii) both the number of uses $K$ and BS
antennas $N$ increase asymptotically by keeping their ratio
$\kappa=\frac{N}{K}$ fixed; and with scaling the power with the
number of antennas $N$, we obtain the corresponding SINRs, in
order to scrutinize their properties. The purpose of this analysis
is to exploit the reduction of the interference and thermal noise
due to the property of orthogonal channels vectors between the BS
and the users as $N \to \infty$ as well as to achieve increase of
the sum-rate due to its dependence of $N$.

\subsubsection{$N \to \infty$ with fixed $p_{r}$ and $K$}
Keeping in mind that an Erlang distribution with shape and scale
parameters given by $N-K+1$ and $\hat{\beta}_{llk}$, respectively,
can be related with the sum of independent normal RVs $W_1[n-1],
W_2[n-1], ..., W_{2\left(N-K+1\right)}[n-1]$, $X_k[n-1]$ as
follows:
\begin{align}\label{eq large N 1}
    X_k[n-1]
    =
        \frac{\hat{\beta}_{llk}}{2}
        \sum_{i=1}^{2\left(N-K+1\right)}
        W_i^2[n-1].
\end{align}

By the substitution of~\eqref{eq large N 1} into~\eqref{eq Prop1
1} as well as by the use of the law of large numbers, the
nominator and the first term of the denominator in \eqref{eq Prop1
1} converge almost surely to $\alpha^2p_{r}\hat{\beta}_{llk}/2$
and $\alpha^2p_{r} C \hat{\beta}_{llk}/2$ as $N \to \infty$, while
the second term of the denominator goes to $0$. As a result, the
deterministic equivalent of the SINR, $\bar{\gamma_k}$, when $N
\to \infty$, is expressed as:
\begin{align} \label{eq large N 2f}
\gamma_k
     \mathop \rightarrow \limits^{\tt a.s.}
          \bar{\gamma_k} = \frac{1}{C}, \quad \text{as
          $N\to\infty$}.
\end{align}
The bounded SINR is expected because it is already known that as
the number of BS antennas tends to infinity,  both the intra-cell
interference and noise are cancelled out, while the inter-cell
interference due to pilot contamination remains.

\subsubsection{$ K, N \to \infty$ with fixed $p_{r}$ and $\kappa={N}/{K}$}
In practice, the number of serving users $K$ in each cell of next
generation systems is  not much less than the number of BS
antennas $N$. In such case, the application of the law of numbers
does not stand because the channel vectors between the BS and the
users are not anymore pairwisely orthogonal. This in turn induces
new properties at the scenario under study, which are going to be
revealed after the following analysis. Basically, we are going to
derive the deterministic approximation $\bar{\gamma}_k$ of the
SINR ${\gamma}_k$ such that
\begin{align}
{\gamma}_k- \bar{\gamma}_k\xrightarrow[ N \rightarrow
\infty]{\mbox{a.s.}} 0,
\end{align}
where  $\xrightarrow[ N \rightarrow \infty]{\mbox{a.s.}}$ denotes
almost sure convergence.

\begin{theorem}
The deterministic equivalent $\bar{\gamma}_k$ of the uplink SINR
between user $k$ and its BS with ZF decoder is given by
\begin{align}\label{eq large k theorem}
\bar{\gamma}_k= \frac{
        \alpha^2\hat{\beta}_{llk}\left(\kappa-1 \right)
        }{
        \alpha^2C\hat{\beta}_{llk}\left(\kappa-1 \right)
+
  \sum_{i=1}^{L} \frac{1}{K}{\tt Tr} \tilde{\B{D}}_{li}
        }.
\end{align}
\begin{proof}
Knowing that $\hat{\B{Y}}_i[n]\sim
\CG{\B{0}}{\tilde{\B{D}}_{li}}$, we can write it  as:
\begin{align}\label{eq Yi}
\hat{\B{Y}}_i[n]&=  \B{a}_{i}^\H \tilde{\B{D}}_{li}^{\frac{1}{2}},
\end{align}
where $\B a_{i}\!\!~\!\sim \CG{\B{0}}{\Id_{K}}$. By
substituting~\eqref{eq large N 1} as well as~\eqref{eq Yi}
into~\eqref{eq Prop1 1}, we have
\begin{align} \label{eq large k}
    \gamma_k
    &=
        \frac{
            \alpha^2p_{r}
            \frac{\hat{\beta}_{llk}}{2}
            \sum_{i=1}^{2\left(N-K+1\right)}
            W_i^2[n-1]
            }{
           \alpha^2p_{r} C
            \frac{\hat{\beta}_{llk}}{2}\!\!
            \sum\limits_{i=1}^{2\left(N-K+1\right)}\!\!
            W_i^2[n-1]
            +p_r
            \sum\limits_{i =1}^{L}
                    \B{a}_i^H
                    \tilde{\B{D}}_{li}
                    \B{a}_i
        +
            1
            }.
\end{align}
Next, if we divide both the nominator and denominator of \eqref{eq
large k} by $2\left( N-K+1 \right)$ and
by using \cite[Lemma~1]{Truong}, under the assumption that
$\tilde{\B{D}}_{li}$ has uniformly bounded spectral norm with
respect to $K$, we arrive at the desired result \eqref{eq large k
theorem}.
\end{proof}
\end{theorem}

\begin{remark}
Interestingly, in contrast to~\eqref{eq large N 2f}, the SINR is
now affected by intra-cell interference  as well as inter-cell
interference and it is independent of the transmit power. In fact,
the former justifies the latter, since both the desired and
interference signals are changed by the same factor, if each user
changes its power. Note that the interference terms remain because
they depend on both $N$ and $K$; however, the dependence of
thermal noise only from $N$ makes it vanish. As
expected,~\eqref{eq large k theorem} coincides with~\eqref{eq
large N 2}, if $N\gg K$, i.e., when $\kappa \to \infty$, the SINR
goes asymptotically to ${1}/{C}$.
\end{remark}

Next, the deterministic equivalent sum-rate can be obtained by
means of the dominated convergence~\cite{Billingsley} and the
continuous mapping theorem~\cite{Vaart} as
\begin{align}
 R_{lk}(p_r,\alpha)- \log_2\left( 1+ \bar{\gamma}_k\right) \xrightarrow[ N \rightarrow \infty]{\mbox{a.s.}} 0.
\end{align}

\subsubsection{Power-Scaling Law}

Let consider $p_{r} = E/\sqrt{N}$, where $E$ is fixed regardless
of $N$. Given that  $\hat{\beta}_{llk}$ depends on
$p_{\mathrm{tr}}=\frac{\tau E}{\sqrt{N}}$, we have from~\eqref{eq large N 2f}
that for fixed $K$ and $N\to\infty$,
\begin{align} \label{eq large N 2}
\gamma_k \mathop \rightarrow \limits^{\tt a.s.}
\frac{\alpha^{2}\tau E^2 {\beta}_{llk}^2}{\alpha^{2} \tau E^2 C
{\beta}^2_{llk}+1},
\end{align}
which is a non-zero constant. This implies that, we can reduce the
transmit power proportionally to $1/\sqrt{N}$, while remaining a
given quality-of-service. In the case where the BS has perfect CSI
and where there is no relative movement of the users, the result
\eqref{eq large N 2} is identical with the result in
\cite{Mathaiou:ZF_receivers}.

\begin{figure}[t]
    \centering
    \centerline{\includegraphics[width=0.70\textwidth]{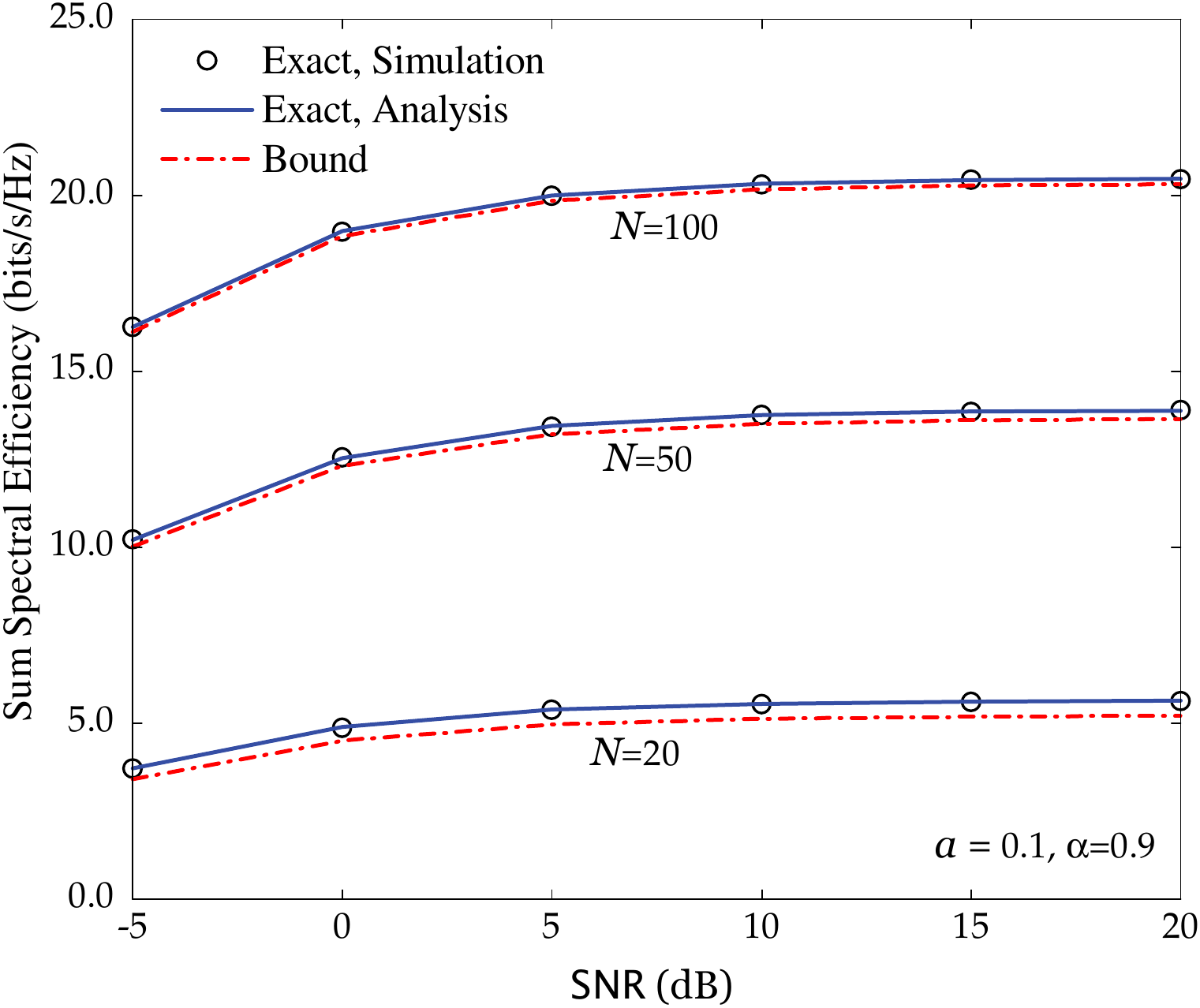}}
    \caption{Sum spectral efficiency  versus $\mathsf{SNR}$ for different $N(a=0.1$ and $\alpha=0.9)$.}
    \label{fig:1}
\end{figure}

\begin{figure}[t]
    \centering
    \centerline{\includegraphics[width=0.70\textwidth]{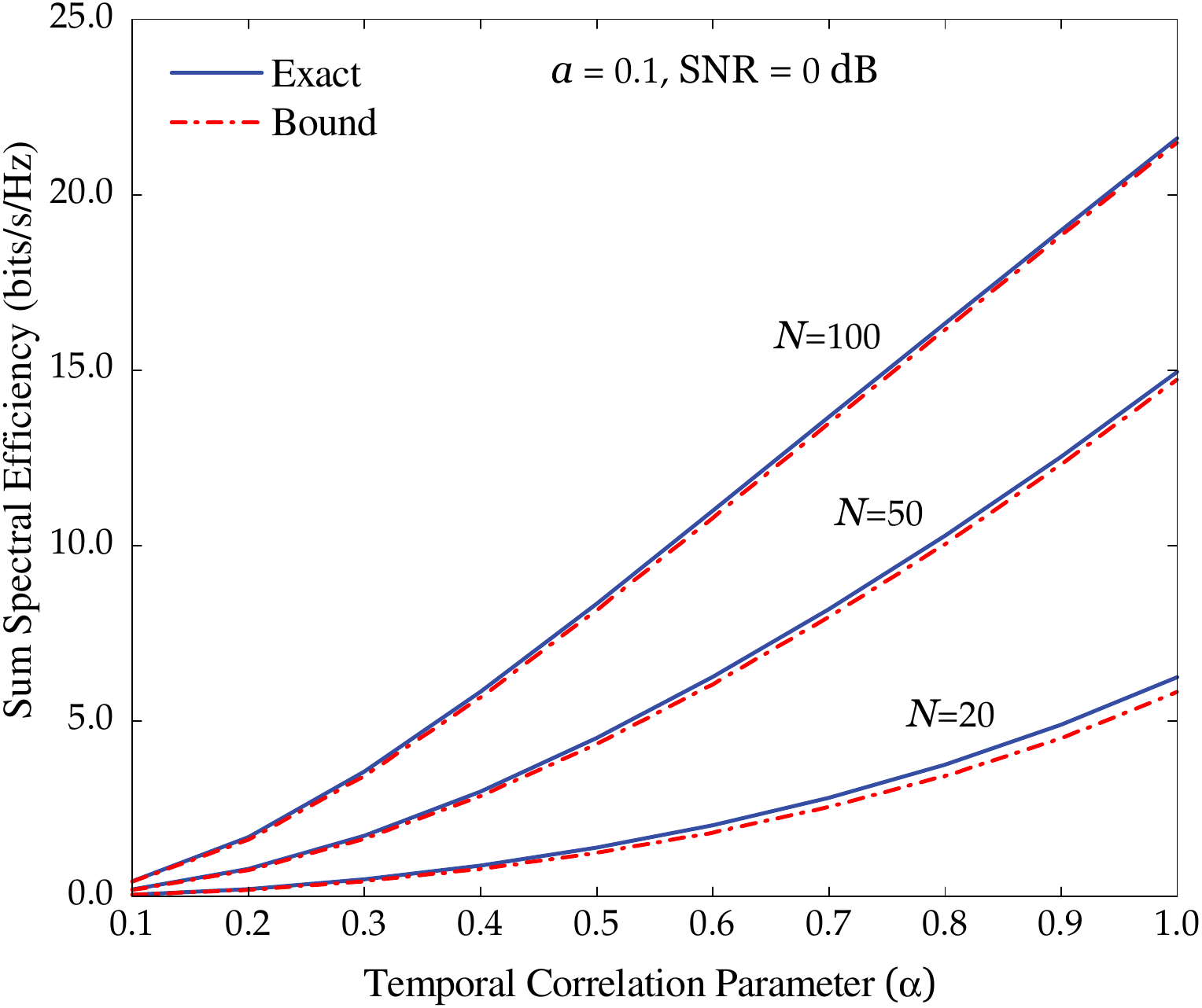}}
    \caption{Sum spectral efficiency  versus $\alpha$ for different $N(a=0.1$ and $\mathsf{SNR}=0\,$dB).}
    \label{fig:2}
\end{figure}

\begin{figure}[t]
    \centering
    \centerline{\includegraphics[width=0.70\textwidth]{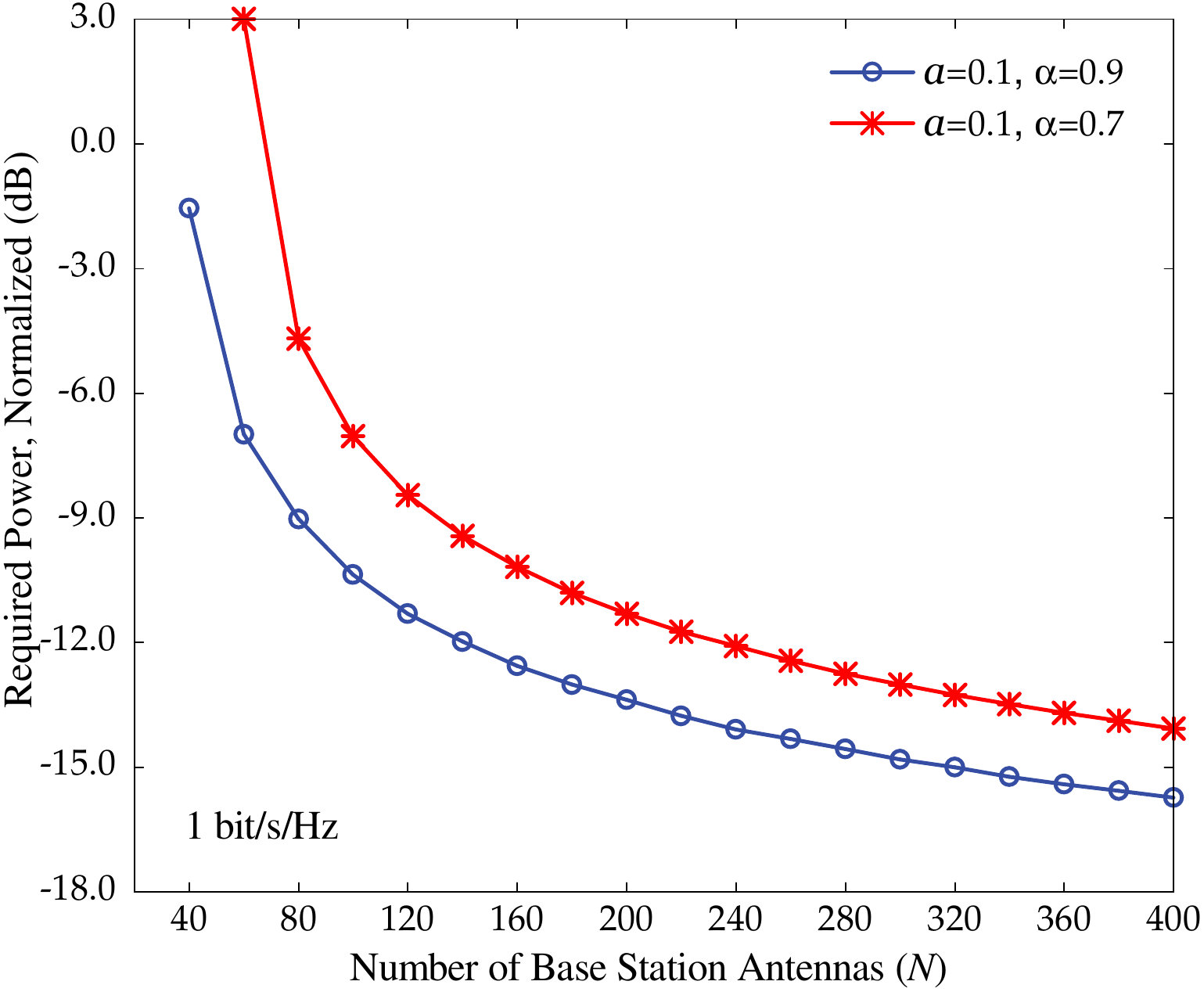}}
    \caption{Transmit power required to achieve 1 bit/s/Hz per user versus $N(a=0.1$, $\alpha=0.7$ and $\alpha=0.9)$.}
    \label{fig:3}
\end{figure}

\begin{figure}[t]
    \centering
    \centerline{\includegraphics[width=0.72\textwidth]{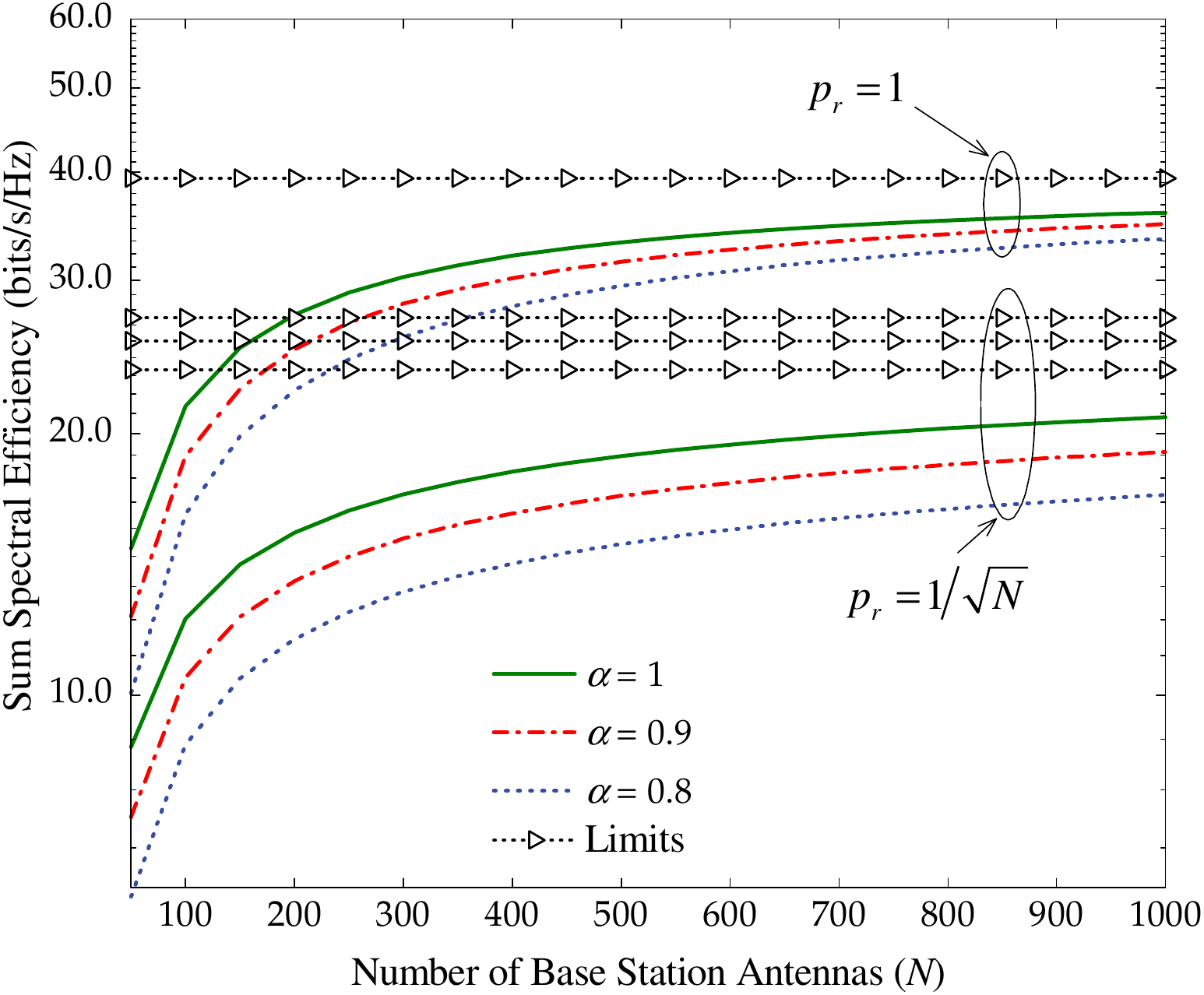}}
    \caption{Sum spectral efficiency  versus $N$ for different $\alpha$.}
    \label{fig:4}
\end{figure}

\begin{figure}[t]
    \centering
    \centerline{\includegraphics[width=0.70\textwidth]{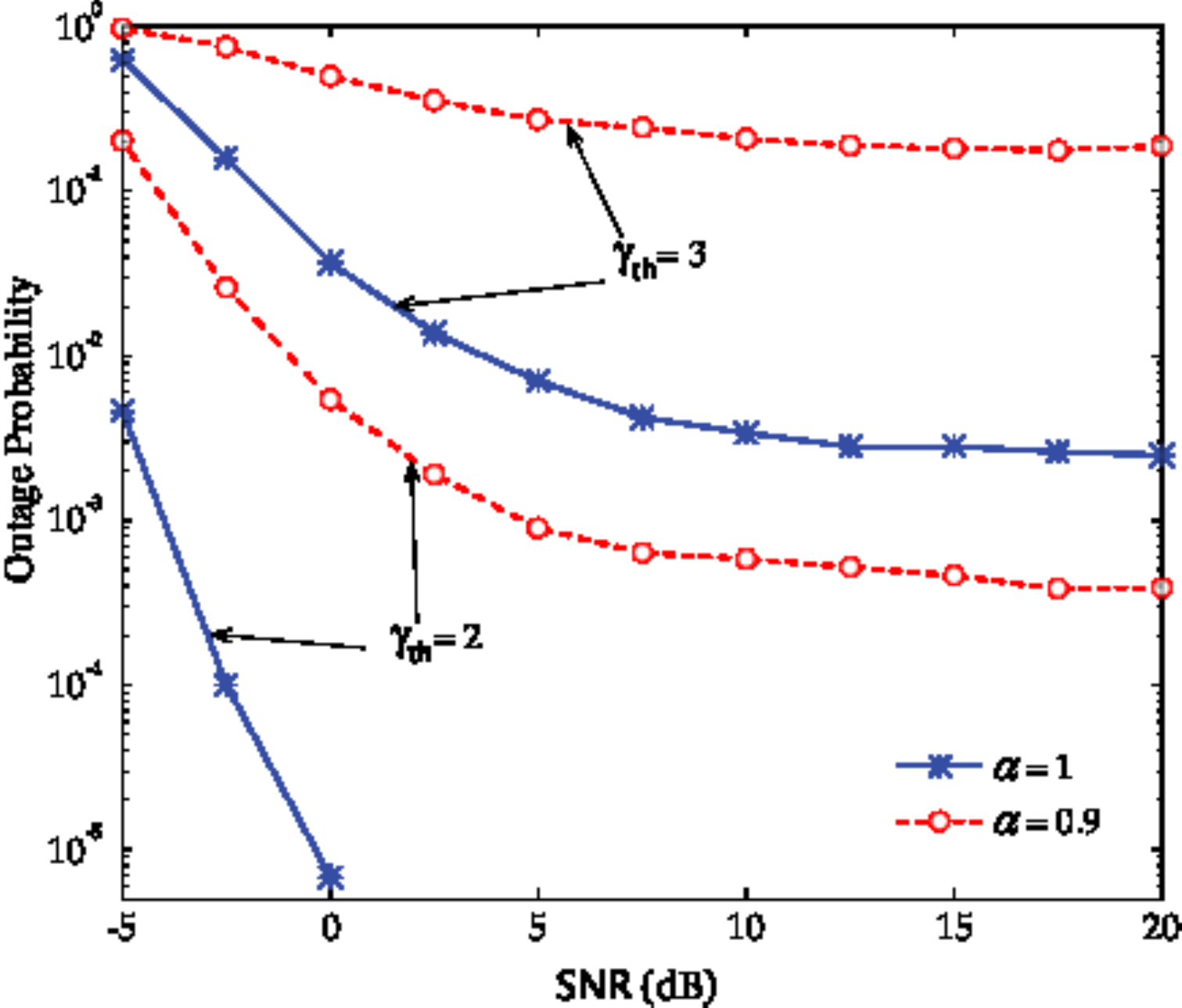}}
    \caption{Outage probability  versus $\mathsf{SNR}$ for different $\alpha$ and $\gamma_{\tt th}$ ($N=100$).}
    \label{fig:5}
\end{figure}

\section{Numerical Results}
In this section, we present numerical results to verify our
analysis by considering a cellular network with $L=7$ cells and
$K=10$ users per cell. The coherence interval is $T=200$ symbols
(which corresponds to a coherence bandwidth of 200 kHz and a
coherence time of 1 ms) and the length of training duration is
$\tau=K$ symbols. Regarding the large-scale coefficients
$\beta_{lik}$, we assume a simple scenario: $\beta_{llk}=1$ and
$\beta_{lik}=a$, for $k=1, \ldots, K$, and $i\neq l$. Note that
$a$ can be considered as an intercell interference factor. In all
examples, we choose $a=0.1$. Furthermore, we define
$\mathsf{SNR}\triangleq p_{r}$.

In the following, we will examine the sum spectral efficiency
which is defined as:
\begin{align} \label{eq num 1}
\mathcal{S}_l \triangleq \left(1-\frac{\tau}{T}\right)\sum_{k=1}^K
 R_{lk}\left(p_r,\alpha \right),
\end{align}
where $ R_{lk}\left(p_r,\alpha \right)$ is given by \eqref{Rate
prop}.

Figure~\ref{fig:1} shows the sum spectral efficiency versus the
SNR for $N=20$, $50$, and $100$, at the intercell interference
factor $a=0.1$, and the temporal correlation parameter
$\alpha=0.9$. The ``Exact, Analysis'' curves are computed by using
\eqref{Rate prop}, the ``Exact, Simulation'' curves are generated via
\eqref{eq: SINR} using Monte-Carlo simulations, while the ``Bound'' curves are
obtained by using the bound result given in Proposition~\ref{Prop BoundRate 1}. The
exact match between the analytical and simulated results validates
our analysis. It can be seen from the figure that, the proposed
bound is very tight, especially for large antenna arrays.
Furthermore, we can see that, at high SNR, the sum spectral
efficiency saturates. This is due to the fact that when SNR
increases, both the desired signal power and intercell
interference power are increased. To improve the system
performance, we can use more antennas at the BS. At
$\mathsf{SNR}=5$dB, the sum spectral efficiencies can be increased
by the factors of $2.5$ or $5.5$ if we increase $N$ from $20$ to
$50$ or from $20$ to $100$, respectively.

Next, we examine the effect of the temporal correlation parameter
on the system performance as well as the tightness of our proposed
bound given in Proposition~\ref{Prop BoundRate 1}. Figure~\ref{fig:2} presents the sum
spectral efficiency as a function of the temporal correlation
parameter $\alpha$, at $\mathsf{SNR}=0$dB, for $N=20, 50$, and
$100$. We can see that the system performance degrades
significantly when the temporal correlation parameter decreases
(or the time variation in the channel increases). Half of the
spectral efficiency is reduced when $\alpha$ reduces from 1 to
$0.6$. Furthermore, at low $\alpha$, using more antennas at the BS
does not help improve the system performance much. Regarding the
tightness of the proposed bound, we can see that the bound is very
tight across the entire temporal correlation range.

Figure~\ref{fig:3} shows the transmit power, $p_{r}$, that is
needed to reach $1$ bit/s/Hz per user. Here, we choose $a=0.1$ and
$\alpha=0.7$ or $0.9$. As expected, the required transmit power
decreases significantly when we increase the number of BS
antennas. By doubling the number of BS antennas, we can cut back
the transmit power by approximately $1.5$dB. This observation is
in line with the results of~\cite{Ngo_Energy}.

To further verify our analysis on large antenna limits, we
consider Figure~\ref{fig:4}. Figure~\ref{fig:4} shows the sum
spectral efficiency versus the number of base station antennas for
different values of $\alpha$, and for two cases: the transmit
power, $p_{r}$, is fixed regardless of $N$, and the transmit power
is scaled as $p_{r}=1/\sqrt{N}$. The ``Limits" curves are computed
via the results obtained in Section~III-C. As expected, as the
number of the base station antennas increases, the sum spectral
efficiencies converge to their limits. When the transmit power is
fixed, the asymptotic performance (as $N\to\infty$) does not
depend on the temporal correlation parameter. By contrast, when
the transmit power is scaled as $1/\sqrt{N}$, the asymptotic
performance depends on $\alpha$.

Finally, we consider the outage performance versus SNR at $N=100$,
for different temporal correlation parameters ($\alpha=1$, and
$0.9$), and for different threshold values ($\gamma_{\tt th}=2$,
and $3$). See Figure~\ref{fig:5}. We can see that, the outage
probability strongly depends on $\alpha$. At $\mathsf{SNR}=0$ dB,
by reducing $\alpha$ form 1 to $0.9$, the outage probability
increases from $7\times 10^{-6}$ to $5\times 10^{-3}$, and from
$3\times 10^{-2}$ to $5\times 10^{-1}$ for $\gamma_{\tt th}=2$,
and $3$, respectively. In addition, the outage probability
significantly improves when the threshold values are slightly
reduced. This is due to the fact that, with large antenna arrays,
the channel hardening occurs, and hence, the SINR concentrates
around its mean. As a results, by slightly reducing the threshold
values, we can obtain a very low outage probability.

\section{Conclusions}
In this paper, we characterized the uplink performance of a
cellular  network taking into account both the well-known pilot
contamination and the unavoidable, but less studied, time
variation. The latter effect, inherent in the vast majority of
propagation scenarios, stems from the user mobility. Summarizing
the main contributions of this work,  new analytical closed-form
expressions for the PDF of the SINR and the corresponding
achievable sum-rate, that hold for any finite number of antennas,
were derived. Moreover, a complete investigation of the low-SNR
regime took place. Neveretheless, asymptotic expressions in the
large antenna/user limit were also obtained, as well as the
power-scaling law was studied. As a final point, numerical
illustrations depicted how the time variation affects the
performance in various Doppler shifts for finite and infinite
number of antennas. Notably, the outcome is that large number of
antennas should be preferred even in time-varying conditions.

\appendix

\subsection{Proof of Theorem~\ref{Theo 1}} \label{sec:proof:prop
rate1}
The uplink ergodic rate from the $k$th user in the $l$th
cell to its BS (in \emph{bits/s/Hz}) is
given by
%
%
\begin{align}
     &R_{lk}\!\left(\! p_{r},\alpha\! \right)
    \!=\!
        \EXs{X_k, Y_k\!\!\!}{
            \!\log_2\!\!
            \left(\!\!
                1\!+\!
                \frac{
                    p_{r} \alpha^2 X_k[n-1]
                    }{
                    p_{r} \alpha^2 C X_k[n\!-\!1] \!+\! p_{r} Y_k[n]\!+\! 1
                }
            \!\right)
        } \nonumber
    \\
 &\!=\!
    \int_0^{\infty}\!\!\!\!
    \int_0^{\infty}\!\!\!
           \log_2
            \left(\!
                1\!+\!
                \frac{
                    p_{r} \alpha^2 x
                    }{
                    p_{r} \alpha^2 C x + p_{r} y+ 1
                }
            \!\right)\!
            \PDF{X_k}{x}
            \PDF{Y_k}{y}
    dx dy
    \nonumber
    \\
    &=
        \sum_{p=1}^{\varrho\left( \mathbf{\mathcal{A}}_k\right)}
        \sum_{q=1}^{\tau_p \left( \mathbf{\mathcal{A}}_k\right)}\!
            \frac{
                 \mathcal{X}_{p,q} \left( \mathbf{\mathcal{A}}_k\right)
               \mu_{k,p}^{-q} \log_2 e
            }{
                \left(
                    q\!-\!1
                \right)
        !
                \left(N\!-\!K\right)!
                \hat{\beta}_{llk}^{N-K+1}
            }
        \nonumber
       \\
       &  \!\times\!
    \!\int_0^{\infty}\!\!\!
    \!\int_0^{\infty}\!\!
            \ln\!\!
            \left(\!\!
                1\!+\!
                \frac{
                    p_{r} \alpha^2 x
                    }{
                    p_{r} \alpha^2 C x \!+\! p_{r} y \!+ \!1
                }
            \!\right)\!
            x^{N\!-\!K} e^{\frac{-x}{\hat{\beta}_{llk}}}
            y^{q-1}
           e^{\frac{-y}{\mu_{k,p}}}      dx dy\nn\\
           &= \sum_{p=1}^{\varrho\left( \mathbf{\mathcal{A}}_k\right)}
        \sum_{q=1}^{\tau_p \left( \mathbf{\mathcal{A}}_k\right)}\!
            \frac{
                 \mathcal{X}_{p,q} \left( \mathbf{\mathcal{A}}_k\right)
               \mu_{k,p}^{-q} \log_2 e
            }{
                \left(
                    q\!-\!1
                \right)
        !
                \left(N\!-\!K\right)!
                \hat{\beta}_{llk}^{N-K+1}
            }
        \nonumber
       \\
       &  \!\times\! \bigg(
    \!\underbrace{\int_0^{\infty}\!\!\!\!
    \!\int_0^{\infty}\!
            \ln\!\!
            \left(\!
                1\!+\!
                \frac{
                    p_{r} \alpha^2 \left(\! C\!+\!1 \!\right)x
                    }{
                     p_{r} y \!+ \!1
                }
            \!\right)\!
            x^{N\!-\!K} e^{\frac{-x}{\hat{\beta}_{llk}}}
            y^{q-1}
           e^{\frac{-y}{\mu_{k,p}}}      dx dy}_{\triangleq \mathcal{I}_1} \nn\\
           &\!-\!
           \!\underbrace{\int_0^{\infty}\!
    \!\int_0^{\infty}\!
            \ln
            \left(\!
                1\!+\!
                \frac{
                    p_{r} \alpha^2 C x
                    }{
                   \ p_{r} y \!+ \!1
                }
            \!\right)\!
            x^{N\!-\!K} e^{\frac{-x}{\hat{\beta}_{llk}}}
            y^{q-1}
           e^{\frac{-y}{\mu_{k,p}}}dx dy}_{\triangleq \mathcal{I}_2}\!\!\bigg)\label{Rate 1_intermediate}\!\!\!\! \\
           &=
\sum_{p=1}^{\varrho\left( \mathbf{\mathcal{A}}_k\right)}
        \sum_{q=1}^{\tau_p \left( \mathbf{\mathcal{A}}_k\right)}\!
            \frac{
                 \mathcal{X}_{p,q} \left( \mathbf{\mathcal{A}}_k\right)
               \mu_{k,p}^{-q} \log_2 e
            }{
                \left(
                    q\!-\!1
                \right)
        !
                \left(N\!-\!K\right)!
                \hat{\beta}_{llk}^{N-K+1}
            }    \left(  \mathcal{I}_{1}-\mathcal{I}_{2} \right).\label{Rate 1}
\end{align}

We first derive $\mathcal{I}_{1}$ by evaluating the integral over
$x$. By using \cite[Eq.~(4.337.5)]{GR:07:Book}, we obtain
\begin{align} \label{Rate 2}
  \mathcal{I}_{1}
    &=   \sum_{t=0}^{N-K} \!\int_0^{\infty}\!
        \left[
-f(y)^{N-K-t}
                e^{-f(y)}
                \mathrm{Ei}\left(f(y) \right)
        \right.
        \nonumber
        \\
        &
            \left.
                +
                \sum_{u=1}^{N-K-t}
                \left(u-1 \right)!
                f(y)^{N-K-t-u}
            \right]
           y^{q-1}
           e^{\frac{-y}{\mu_{k,p}}}
    dy,
\end{align}
where $f(y)\triangleq -\frac{ p_{r}  y +1}{\hat{\beta}_{llk} p_{r}
\alpha^2  \left( C+1 \right)}$. Using \cite[Lemma~1]{Duong} and
\cite[Eq.~(39)]{KA:06:WCOM}, we can easily obtain
$\mathcal{I}_{1}$ as given in \eqref{eq Prop2 1a}. Similarly, we
obtain $\mathcal{I}_{2}$ as given in \eqref{eq Prop2 1b}.
Substitution of  $\mathcal{I}_{1}$ and $\mathcal{I}_{2}$
into~\eqref{Rate 1} concludes the proof.

\subsection{Proof of Proposition~\ref{Prop BoundRate
1}}\label{propBound Rate}

By using  Jensen's inequality, we have
\begin{align} \label{Rate 1b}
      R_{lk}\left( p_{r},\alpha \right)
&=\EX{\log_2\left(1+ \gamma_{k}\right)}
=\EX{\log_2\left(1+ \frac{1}{1/\gamma_{k}}\right)}\nn\\
&\ge \log_2\left(1+ \frac{1}{\EX{1/\gamma_{k}}}\right) \triangleq
R_{L}\left( p_{r},\alpha \right).
\end{align}
To compute $R_{L}\left( p_{r},\alpha \right)$, we need to compute
$\EXs{}{1/\gamma_{k}}$. From \eqref{eq Rate 1}, we have
\begin{align}
 &\EXs{}{\frac{1}{\gamma_{k}}}=C+\frac{1}{\alpha^{2}}\sum_{i=1}^{L}\EXs{}{\bigg\|\left[\hat{\B{G}}_{ll}^{\dagger}[n-1]\right]_k \tilde{\B{E}}_{li}[n]\bigg\|^{2}}\nn\\
 &\hspace{3.5cm}+\frac{1}{\alpha^{2}p_{r}}\EXs{}{\bigg\|\left[\hat{\B{G}}_{ll}^{\dagger}[n-1]\right]_k \bigg\|^{2}}\nn\\
  &\!=\!C\!+\!\frac{1}{\alpha^{2}}\!\! \EXs{}{\bigg\|\!\!\left[\!\hat{\B{G}}_{ll}^{\dagger}[n\!-\!1]\!\right]_k \!\bigg\|^{2}}\!\!\left(\sum_{i=1}^{L}\!\sum_{k=1}^{K}\! \left(\! \beta_{lik}\!-\!\alpha^{2}\hat{\beta}_{lik} \!\right)\!+\!\frac{1}{p_{r}} \!\right)\nn\\
  &\!=\!C\!+\!\frac{1}{\left(\! N\!-\!K \!\right)\! \alpha^{2}\hat{ \beta}_{llk}}\left(\sum_{i=1}^{L}\sum_{k=1}^{K}\! \left( \beta_{lik}\!-\!\alpha^{2}\hat{\beta}_{lik} \!\right)\!+\!\frac{1}{p_{r}} \right).\label{Rate 1d}
 \end{align}
 In the third equality of \eqref{Rate 1d}, we have considered the independence between the two variables,
while in the last equality, we have used the following result:
 \begin{align}
  &\EXs{}{\bigg\|\left[\hat{\B{G}}_{ll}^{\dagger}[n-1]\right]_k \bigg\|^{2}}=\EXs{X_{k}}{\frac{1}{X_{k}[n-1]}}\nn\\
  &\!=\! \int_{0}^{\infty}\!\!\!\frac{
            e^{-x/\hat{ \beta}_{llk}}
            }{
            \left(N-K\right)!
           \hat{ \beta}_{llk}^{2}
            }\!\!
        \left(
            \frac{
                x
                }{
               \hat{ \beta}_{llk}
                }
        \right)^{N\!-\!K\!-\!1}\!\!\!\mathrm{d}x= \frac{1}{\left( N-K \right) \hat{ \beta}_{llk}}.\label{Rate 1e}
 \end{align}
Note that we have used~\cite[Eq.~(3.326.2)]{GR:07:Book} to obtain
\eqref{Rate 1e}. Thus, the desired result \eqref{eq: LB} is
obtained from~\eqref{Rate 1b} and ~\eqref{Rate 1d}.

\subsection{Proof of Theorem~\ref{theo 2}}  \label{proof:Prop out}

Clearly, from \eqref{eq Prop1 1}, $\gamma_{k} < 1/C$. Thus, if
$\gamma_{{\tt th}} \geq 1/C$, then  $P_{{\tt out}}\left(
\gamma_{{\tt th}}\right) =1$. Next, we consider the case where
$\gamma_{{\tt th}} < 1/C$. Taking the probability of the
instantaneous SINR $\gamma_{k}$, given by~\eqref{eq Prop1 1}, we
can determine the outage probability as
\begin{align}
 \!\!\!\!\!\!&P_{{\tt out}}
 \!= \!{\tt Pr}\left( \frac{\alpha^2 p_{r} X_k}{\alpha^2 p_{r} C X_k + p_{r} Y_k+ 1}\leq\gamma_{{\tt
 th}}\right)\nn\\
 &\!=\! \int_0^\infty {\tt Pr}\left( X_k< \frac{\gamma_{{\tt
th}}\left( p_{r} Y_k +1 \right)}{ \alpha^{2}p_{r} - \gamma_{{\tt
 th}} \alpha^{2} p_{r} C}\left.\right|Y_k \right) p_{Y_k}(y)\mathrm{d}y\nn\\
  &\!=\!1\!-\!e^{\frac{-\gamma_{{\tt th}}}{p_{r}\bar{\gamma}_{{\tt th}} }}\!
\sum^{N-K}_{t=0}\!\! \int_{0}^{\infty}\! \!\!e^{
\frac{-y}{\bar{\gamma}_{{\tt th}} }}
 \!\sum^{N-K}_{t=0}\!\!\frac{\left( \frac{ \gamma_{{\tt th}}}{\bar{\gamma}_{{\tt th}}} \right)^{\!\!t}}{t!}\!\left( \!y\!+\!\frac{1}{p_{r}} \!\right)^{\!\!t}\!\!p_{Y_k}(y)\mathrm{d}y \nn\\
 &\!=\!1\!-\!e^{\frac{-\gamma_{{\tt th}}}{p_{r}\bar{\gamma}_{{\tt th}} }}
\sum_{p=1}^{\varrho\left( \mathbf{\mathcal{A}}_k\right)}
\sum_{q=1}^{\tau_p \left( \mathbf{\mathcal{A}}_k\right)}
 \sum_{t=0}^{N-K}
 \mathcal{X}_{p,q} \left(
 \mathbf{\mathcal{A}}_k\right)\frac{\mu_{k,p}^{-q}}{\left(q-1\right)} \frac{\left( \frac{ \gamma_{{\tt th}}}{\bar{\gamma}_{{\tt th}}} \right)^{t}}{t!}\nn\\
 &\hspace{3cm}\times\int_{0}^{\infty} y^{q-1}e^{\frac{-y}{\bar{\gamma}_{{\tt th}})} }
 \left( y+\frac{1}{p_{r}} \right)^{t}\mathrm{d}y \nn\\
 &\!=\!1\!-\!e^{\frac{-\gamma_{{\tt th}}}{p_r\bar{\gamma}_{{\tt th}}}}\!\!
\sum_{p=1}^{\varrho\left( \mathbf{\mathcal{A}}_k\right)}\!
\sum_{q=1}^{\tau_p \left( \mathbf{\mathcal{A}}_k\right)}\!
 \sum_{t=0}^{N-K} \!\!
  \sum_{s=0}^{t}\!\! \binom{t}{s}
 \mathcal{X}_{p,q} \!\left(
 \mathbf{\mathcal{A}}_k\!\right)\!\frac{\Gamma\left( s\!+\!q \right)\! \bar{\gamma}_{{\tt
 th}}^{s+q}}{\mu_{k,p}^{q}\left(q\!-\!1\right)},
 \label{eq:CDF3}
\end{align}
where $\bar{\gamma}_{{\tt th}}\triangleq \hat{\beta}_{llk}\left(
\alpha^2 -\alpha^{2}  C \gamma_{{\tt th}} \right)$, and where in
the third equality, we have used that the cumulative density
function of $X_k$ (Erlang variable) is
\begin{align}
F_{X_k}(x)&={\tt Pr}\left(X_k\leq x\right)\nn\\
&=1-\exp\left(-\frac{x}{\hat{\beta}_{llk}}\right)\sum_{t=0}^{N-K}\frac{1}{t!}\left(\frac{x}{\hat{\beta}_{llk}}\right)^t
\label{eq:CDF4}.
\end{align}
The last equality of \eqref{eq:CDF3} was derived after applying
the binomial expansion of $(y+1/p_r)^t$
and~\cite[Eq.~(3.351.1)]{GR:07:Book}.

\setcounter{eqnback}{\value{equation}} \setcounter{equation}{55}
\begin{figure*}[!t]
\begin{small}
\begin{align}
 \!\!\!\!\ddot{R}_{lk}\left( p_{r}, \alpha \right)=\frac{1}{\mathrm{ln}2} \EXs{X_k, Y_k}{\frac{\alpha^{2 }X_k[n-1]\left( \alpha^{4}X_k^{2}[n-1]+2\varsigma_k\left( 1+p_{r} \varsigma_k \right)^{2} \left( 1+\alpha^{2}p_{r} X_k[n-1] +p_{r} \varsigma_k\right)\right)}{\left( \alpha^{4}p_{r}^{2}\left(C+1  \right)X_k^{2}[n-1]+p Y_k[n]+1 \right)^{2}\left( \alpha^{4}p_{r}^{2}C X_k^{2}[n-1]+p_{r} Y_k[n]+1 \right)^{4}}},\label{eq: second_derivative}
\end{align}
\end{small}
\hrulefill
\end{figure*}
\setcounter{eqncnt}{\value{equation}}
\setcounter{equation}{\value{eqnback}}
\subsection{Proof of Theorem~\ref{Theo 3}} \label{sec:low snr} The
initial step for the derivation of the minimum transmit energy per
information bit is to cover the need for exact expressions
regarding the  derivatives of  ${R}_{lk}\left( p_{r}, \alpha
\right)$. In particular, this can be given by
\begin{align}\label{eq: first_derivative}
\!&\!\!\dot{R}_{lk}\left(\! p_{r}, \alpha \!\right)\!=\!
\frac{1}{\mathrm{ln}2}\nn\\
\!&\!\!\!\!\times\!\EXs{X_k, Y_k\!\!\!}{\!\!\frac{\alpha^{2 }X_k[n\!-\!1]\big/\!\!\left(\!
\alpha^{4}p_{r}^{2}C X_k^{2}[n\!-\!1]\!+\!p_{r} Y_k[n]\!+\!1
\!\right)\!\!}{\left(\! \alpha^{4}p_{r}^{2}\left(\!C\!+\!1
\!\right)\!X_k^{2}[n-1]\!+\!p_{r} Y_k[n]+1 \!\right)}}\!\!.
\end{align}
Easily, its value at $p_{r}=0$ is
\begin{align}\label{eq: first_derivative at zero}
 \dot{R}_{lk}\left( 0, \alpha \right)= \frac{1}{\mathrm{ln}2}\EXs{X_k}{\alpha^{2 }X_k[n-1]}.
\end{align}
Taking into account that $X_k[n-1]$ is Erlang distributed, its
expectation can be written as
\begin{align}\label{eq: mean_value}
 \EXs{X_k}{X_k[n-1]}= \left( N-K+1 \right)\hat{\beta}_{llk}.
\end{align}
Substituting~\eqref{eq: mean_value} and~\eqref{eq:
first_derivative at zero} into~\eqref{eq: minimum energy}, we
obtain the desired result.

The second derivative of ${R}_{lk}\left( p_{r}, \alpha \right)$,
needed for the evaluation of the wideband slope, is given by
\eqref{eq: second_derivative} shown at the top of the previous
page, where $\varsigma_k \triangleq \alpha^{2}C X_k[n-1]+Y_k[n]$.
Hence, $\ddot{R}_{lk}\left( 0, \alpha \right)$ can be expressed
by
\setcounter{equation}{56}\begin{align}\label{eq: second_derivative
at zero} \ddot{R}_{lk}\left( 0, \alpha \right)
&=\frac{1}{\mathrm{ln}2}\mathbb{E}_{X_k, Y_k}\left\{
\alpha^{6}X_k^{3}[n-1]+2 \alpha^{4}C X_k^{2}[n-1]\right.
\nn\\
&\hspace{2.5cm}\left.+2 \alpha^{2}X_k[n-1]Y_k[n]\right\}.
\end{align}
The moments of $X_k[n-1]$ are obtained by means of the
corresponding derivatives of its moment generating function (MGF)
at zero $\mathrm{M}_{X_{k}}^{\left( n \right)}\left( 0 \right)$,
i.e., $\EXs{X_k}{X_k^{n}[n-1]}=\mathrm{M}_{X_{k}}^{\left( n
\right)}\left( 0 \right)$. Thus, having in mind that the MGF of
the Erlang distribution is
\begin{align}\label{eq: MGF}
\mathrm{M}_{X_{k}}\left(t  \right)=\frac{1}{\left( 1-
\hat{\beta}_{llk}t\right)^{N-K+1}},
\end{align}
we can obtain the required moments of $X_k[n-1]$ as
\begin{align}\label{eq: 1_MGF}
 \EXs{X_k}{X_k^{2}[n-1]}&=\mathrm{M}_{X_{k}}^{\left( 2 \right)}\left( 0 \right)\nn\\
 &=\frac{\Gamma\left( N-K+3 \right)}{\Gamma\left( N-K+1 \right)}\hat{\beta}_{llk}^{2}\\
 \EXs{X_k}{X_k^{3}[n-1]}&=\mathrm{M}_{X_{k}}^{\left( 3 \right)}\left( 0 \right)\nn\\
 &=\frac{\Gamma\left( N-K+4 \right)}{\Gamma\left( N-K+1 \right)}\hat{\beta}_{llk}^{3}.\label{eq: 2_MGF}
\end{align}

In addition, since $X_k[n-1]$ and $Y_k[n]$ are uncorrelated, we
have  $\EXs{X_k,Y_{k}}{ X_k[n-1]Y_k[n]}=\EXs{X_k}{
X_k[n-1]}\EXs{Y_k}{Y_k[n]}$. In other words, it is necessary to
find the expectation of $Y_k[n]$. As aforementioned, the PDF of
$Y_k[n]$ obeys~\eqref{eq PDF1 3} and has expectation given by
definition as
\begin{align}
  \EXs{Y_k}{Y_k[n]}&=\int_{0}^{\infty}y\PDF{Y_k}{y}\mathrm{d}y\nn\\
    &=
        \sum_{p=1}^{\varrho\left( \mathbf{\mathcal{A}}_k\right)}
        \sum_{q=1}^{\tau_p \left( \mathbf{\mathcal{A}}_k\right)} \!\!\!
            \mathcal{X}_{p,q} \left( \mathbf{\mathcal{A}}_k\right)
            \frac{
                \mu_{k,p}^{-q}
            }{
                \left(
                    q-1
                \right)
                !
            }
           \int_{0}^{\infty}\!\!\! y^{q}
            e^{\frac{-y}{\mu_{k,p}}}\mathrm{d}y\nn\\
        &= \sum_{p=1}^{\varrho\left( \mathbf{\mathcal{A}}_k\right)}
        \sum_{q=1}^{\tau_p \left( \mathbf{\mathcal{A}}_k\right)} \!
            \mathcal{X}_{p,q} \left( \mathbf{\mathcal{A}}_k\right)
            \frac{
                \mu_{k,p}^{-q}q
            }{
                \left(
                    q-1
                \right)
                !
            },\label{eq: mean y}
\end{align}
where we have used~\cite[Eq.~(3.326.2)]{GR:07:Book} as well as the
identity $\Gamma\left( q+1 \right)=q!$. As a result,
$\ddot{R}_{lk}\left( 0, \alpha \right)$ follows by means
of~\eqref{eq: 1_MGF},~\eqref{eq: 2_MGF},~\eqref{eq: mean y}.
Finally, substitution of the~\eqref{eq: first_derivative at zero}
and~\eqref{eq: second_derivative at zero} into~\eqref{eq:
wideband_slope} yields the wideband slope.


\end{document}